\documentclass[prc,aps,eqsecnum,floatfix,nofootinbib]{revtex4-1}
\usepackage{ulem}  
\usepackage{dcolumn}
\usepackage{bm}
\usepackage{color}
\usepackage{amssymb}
\usepackage{amsmath}
\usepackage{graphicx}
\usepackage{amsfonts}
\usepackage{slashed}
\usepackage{pstricks}
\usepackage{float}
\usepackage{hyperref}
\usepackage{array}
\usepackage{epsf}
\usepackage{soul}
\usepackage{multirow}
\usepackage{siunitx}
\allowdisplaybreaks{
\begin{document}
\title{Two- and three-nucleon contact interactions and ground-state energies of light- and
medium-mass nuclei}
\author{R.\ Schiavilla$^{\rm a,b}$, L.\ Girlanda$^{\rm c,d}$, A.\ Gnech$^{\rm b,e}$, A.\ Kievsky$^{\rm f}$, A.\ Lovato$^{\rm g,h}$, L.E.\ Marcucci$^{\rm e,f}$, M.\ Piarulli$^{\rm i}$, and M.\ Viviani$^{\rm f}$}
\affiliation{
$^{\rm a}$\mbox{Department of Physics, Old Dominion University, Norfolk, VA 23529, USA}\\
$^{\rm b}$\mbox{Theory Center, Jefferson Lab, Newport News, VA 23606, USA}\\
$^{\rm c}$\mbox{Department of Mathematics and Physics, University of Salento, 73100 Lecce, Italy} \\
$^{\rm d}$\mbox{INFN-Lecce, 73100 Lecce, Italy} \\
$^{\rm e}$\mbox{Department of Physics, University of Pisa, 56127 Pisa, Italy}\\
$^{\rm f}$\mbox{INFN-Pisa, 56127 Pisa, Italy}\\
$^{\rm g}$\mbox{Physics Division, Argonne National Laboratory, Argonne, IL 60439, USA}\\
$^{\rm h}$\mbox{INFN-TIFPA Trento Institute of Fundamental Physics and Applications, 38123 Trento, Italy}\\
$^{\rm i}$\mbox{Department of Physics, Washington University in St.~Louis, St.~Louis, MO 63130, USA}\\
}
\date{\today}
\begin{abstract}
Classes of two-nucleon ($2N$) contact interactions are developed in configuration space at leading
order (LO), next-to-leading order (NLO), and next-to-next-to-next-to-leading order (N3LO) by fitting
the experimental singlet $np$ scattering length and deuteron binding energy at LO, and $np$ and
$pp$ scattering data in the laboratory-energy range 0--15 MeV at NLO and 0--25 MeV at N3LO.
These interactions are regularized by including two Gaussian cutoffs, one for $T\,$=$\,0$ and the
other for $T\,$=$\,1$ channels.  The cutoffs are taken to vary in the ranges $R_0\,$=$(1.5$--2.3) fm
and $R_1\,$=$(1.5$--3.0) fm.  The 780 (1,100) data points up to 15 (25) MeV energy, primarily differential
cross sections, are fitted by the NLO (N3LO) models with a $\chi^2$/datum about 1.7 or less (well
below 1.5), when harder cutoff values are adopted.   As a first application, we report results for the
binding energies of nuclei with mass numbers $A\,$=$\,3$--6 and 16 obtained with selected
LO and NLO $2N$ models both by themselves as well as in combination with a LO three-nucleon ($3N$)
contact interaction.  The latter is characterized by a single low-energy constant that is fixed to
reproduce the experimental $^3$H binding energy.  The inclusion of the $3N$ interaction largely
removes the sensitivity to cutoff variations in the few-nucleon systems and leads to predictions
for the $^3$He and $^4$He binding energies that cluster around 7.8 MeV and 30 MeV, respectively.
However, in $^{16}$O this cutoff sensitivity remains rather strong. 
Finally, predictions at LO only are also reported for medium-mass nuclei with $A\,$=$\,40$, 48, and 90.

\end{abstract}

\index{}\maketitle
\section{Introduction}
\label{sec:intro}

Understanding the interactions among the constituents of atomic nuclei lies at the heart of nuclear physics and is still a subject of intense research.
Since the advent of the Effective Field Theory (EFT) paradigm~\cite{Weinberg:1990rz,Weinberg:1991um} in the early nineties, two-nucleon ($2N$) chiral interactions have been developed up to fourth (N3LO) and, more recently, fifth (N4LO) order in the low-energy expansion~\cite{Ordonez:1992xp,Ordonez:1993tn,vanKolck:1994yi,Ordonez:1995rz,Entem:2003ft,Epelbaum:2004fk,Epelbaum:2014efa,Epelbaum:2014sza,Piarulli:2014bda,Entem:2017gor,Reinert:2017usi}.  These interactions provide an accurate description of $2N$ scattering data up to the pion production threshold, comparable to that obtained
by phenomenological models~\cite{Stoks:1994wp,Wiringa:1994wb,Machleidt:2000ge}.

Implicit in the definition of an EFT is a cutoff scale $\Lambda$ which marks the separation between the domain of applicability and high-energy scales that characterize processes unresolved by the EFT and whose effects are subsumed in the values of low-energy constants (LECs).  An interesting version of EFT is one in which the cutoff is taken to be smaller than the pion mass, that is, the pion mass represents the heavy scale.
In such a regime, pions are integrated out and the theory only consists of contact terms between two or more
nucleons---pionless EFT~\cite{Chen:1999tn,Bedaque:2002mn} ($\slashed{\pi}$EFT).  A natural question to ask is: how well (or how poorly) will low-energy nuclear structure, including binding energies,
charge radii, and magnetic moments, be accounted for by this simpler EFT?  As a first step in our attempt to answer this question, we construct in this
paper coordinate-space $2N$ contact interactions from fits to scattering observables in a limited range of energies. These $2N$ interactions are complemented by a LO three-nucleon ($3N$) contact
interaction, constrained to reproduce the $^3$H binding energy.  A first set of calculations of the
ground-state energies of the hydrogen and helium isotopes, $^6$Li, and $^{16}$O ($^{40}$Ca,
$^{48}$Ca and $^{90}$Zr) is presented with selected models at LO and NLO (LO only).
Results for the same observable with the N3LO models are limited to the $^3$H, $^3$He/$^4$He/$^6$He and $^6$Li nuclei.

In the $2N$ system, $\slashed{\pi}$EFT reduces to the effective range expansion~\cite{Bethe:1949yr}.  Due to the unnaturally large values of the $2N$ scattering lengths, it is convenient, in order to extend the domain of applicability of the theory, to consider the inverse scattering length as a soft
scale~\cite{Kaplan:1998tg,Kaplan:1998we,hammer:2020}.  As a consequence, this EFT corresponds to an expansion around the unitary limit of infinite scattering
length~\cite{Konig:2016utl,Gattobigio:2019omi}.  By introducing a single expansion parameter, the ratio
of the interaction range to the scattering length~\cite{Bedaque:1998kg,Bedaque:1998km,Braaten:2004rn,deltuva:2020},
such a theory accounts for universal phenomena, such as the Efimov effect~\cite{Efimov:1970zz,Efimov:1971zz,Naidon:2016dpf,Kievsky:2016kzb}, in systems of three and more nucleons.

Depending on the renormalization conditions, two low-energy counting schemes can consistently be implemented~\cite{Epelbaum:2017byx}, the Weinberg counting, in which the magnitude of the LECs entering the interaction follows naive dimensional analysis~\cite{Manohar:1983md,Georgi:1992dw}, and the Kaplan, Savage, and Wise (KSW) counting~\cite{Kaplan:1998tg,Kaplan:1998we}, in which their importance is enhanced. In the present paper, we adhere to Weinberg counting (for related work based on KSW counting see Refs.~\cite{Kirscher:2009aj,Lensky:2016djr}). This implies a certain amount of fine tuning of the two leading LECs, which have a direct connection to the unnaturally large values of the singlet and triplet scattering lengths.  As a matter of fact, we are led to introduce two different cutoffs in the $T\,$=$\,0$ and $T\,$=$\,1$ isospin channels, in order to reduce, in the fitting procedure, the correlations induced by such fine tuning.  Following common practice in the construction of $2N$ interaction models from EFT, we choose to perform an implicit renormalization of the LECs, through the fitting of low-energy experimental data.  Had we chosen to fix each one of the two leading LECs to a single observable, like the corresponding scattering length, we would
have obtained a dependence (running) on the associated cutoff (or renormalization point), one from each renormalization condition.
Since cutoff-independence in the description of other observables is to be expected only up to neglected orders, the implicit renormalization procedure is likely to drive the LECs away from the renormalization group running, except around some special value of the cutoff, which needs not be the same for the two leading LECs. It is expected that, when higher and higher orders are included, the optimal cutoff regions will grow until a plateau is realized, and eventually will overlap. We should also mention that at least two independent cutoffs were found to be necessary in order to derive the rules of Weinberg counting from the Wilsonian renormalization group~\cite{Epelbaum:2017tzp}.

The present paper is organized as follows. In Sec.~\ref{sec:potential} the $2N$ contact interaction is introduced up to N3LO\footnote{We denote the various orders in the expansion of the interaction following the usual convention in pionfull EFT, where NLO is $O(Q^2)$ suppressed relative to LO, and N3LO is $O(Q^2)$ suppressed relative to NLO.  Here $Q$ denotes a low-momentum scale.}, and is regularized to obtain its coordinate space representation.  In Sec.~\ref{sec:fits} the associated LECs are determined through an order-by-order fit to $2N$ scattering observables below 15~MeV and
25~MeV laboratory energies at, respectively, NLO and N3LO, and to the deuteron binding energy. In Sec.~\ref{sec:a34} results for the binding energies of $^3$H, $^3$He, $^4$He, $^6$Li, $^{6}$He,
and $^{16}$O are reported for selected models at LO and NLO, and for the binding energies of $^{40}$Ca,
$^{48}$Ca, and $^{90}$Zr with selected LO models only.
The calculations are carried out with hyperspherical-harmonics (HH) methods
in systems with mass number $3\le A \le 6$, and with auxiliary-field diffusion Monte Carlo (AFDMC)
methods in $A\ge 16$.  Finally, a brief summary and
some concluding remarks are given in Sec.~\ref{sec:concl}.

\section{Contact interactions at LO, NLO, and N3LO: a summary}
\label{sec:potential}

The structure of two-nucleon ($2N$) contact interactions at LO, NLO, and N3LO
is well known~\cite{Ordonez:1995rz}; we provide a brief summary here for completeness.  These interactions consist
of charge-independent (CI) terms at LO, NLO and N3LO, and charge-dependent (CD)
ones at NLO and N3LO.  However, in a departure from common practice,
we require the LO interactions to only act in even partial waves.  We explain
the rationale for such a choice in Sec.~\ref{sec:a34} below.

\subsection{Contact interactions in momentum space}
The interactions in momentum space are listed below order by order in the power
counting ($Q$ denotes generically a low-momentum scale).  The momenta
${\bf k}$ and ${\bf K}$ are defined as ${\bf k}\,$=$\,{\bf p}^\prime-{\bf p}$
and ${\bf K}\,$=$\left({\bf p}^\prime+{\bf p}\right)/2$, where ${\bf p}$ and ${\bf p}^\prime$
are the initial and final relative momenta of the two nucleons, and ${\bm \sigma}_i$ and
${\bm \tau}_i$ denote respectively the Pauli spin and isospin operators:
\begin{itemize}
\item{CI terms of LO ($Q^0$):}
\begin{equation}
v_{\rm LO}^{\rm CI}=C_{01} \, P^{\sigma}_0\, P^\tau_1 +C_{10}\,P^{\sigma}_1\, P^\tau_0  \ ,
\end{equation}
where $P^\sigma_0$ ($P^\tau_0$) and $P^\sigma_1$ ($P^\tau_1$) are spin (isospin) projection
operators on pairs with $S$ ($T$) equal to 0 and 1,
\begin{equation}
P^\sigma_0 =\frac{1-\,{\bm \sigma}_1\cdot{\bm\sigma}_2}{4} \ ,\qquad P^\sigma_1 =\frac{3+{\bm \sigma}_1\cdot{\bm \sigma}_2}{4} \ ,
\end{equation}
and similarly for $P^\tau_0$ and $P^{\tau}_1$;
\item{CI term of NLO ($Q^2$):}
\begin{eqnarray}
v_{\rm NLO}^{\rm CI}({\bf k}, {\bf K})&=& C_1\,k^2
+C_2\,k^2\, {\bm \tau}_1\cdot {\bm \tau}_2
+C_3\,k^2\, {\bm \sigma}_1\cdot {\bm \sigma}_2
+\, C_4\,k^2\,{\bm \sigma}_1\cdot {\bm \sigma}_2\,{\bm \tau}_1\cdot {\bm \tau}_2
+C_5 \,S_{12}({\bf k}) \nonumber\\
&&+\, C_6\,S_{12}({\bf k})\,{\bm \tau}_1\cdot {\bm \tau}_2
+i\, C_7\, {\bf S}\cdot \left({\bf K} \times {\bf k}\right) \ ,
\end{eqnarray}
where $S_{12}({\bf k})= 3\,{\bm\sigma}_1 \cdot {\bf k}\,\, {\bm\sigma}_2 \cdot {\bf k}-
k^2\,{\bm\sigma}_1\cdot{\bm \sigma}_2\,$;
\item{CI terms of N3LO ($Q^4$):}
\begin{eqnarray}
v_{\rm N3LO}^{\rm CI}({\bf k}, {\bf K})&=& D_1\,k^4
+D_2\,k^4\, {\bm \tau}_1\cdot {\bm \tau}_2
+D_3\,k^4\, {\bm \sigma}_1\cdot {\bm \sigma}_2
+\, D_4\,k^4\,{\bm \sigma}_1\cdot {\bm \sigma}_2\,{\bm \tau}_1\cdot {\bm \tau}_2
+D_5 \,k^2\,S_{12}({\bf k}) \nonumber\\
&&+\, D_6\,k^2\,S_{12}({\bf k})\,{\bm \tau}_1\cdot {\bm \tau}_2
+i\, D_7\, k^2\,{\bf S}\cdot \left({\bf K} \times {\bf k}\right) 
+i\,D_8\,k^2\,{\bf S}\cdot \left({\bf K}\, \times {\bf k}\right){\bm \tau}_1\cdot {\bm \tau}_2
+D_{9}\left[{\bf S}\cdot \left({\bf K} \times {\bf k}\right)\right]^2\nonumber\\
&&+\,D_{10}\left({\bf K} \times {\bf k}\right)^2
+D_{11}\left({\bf K} \times {\bf k}\right)^2{\bm \sigma}_1\cdot {\bm \sigma}_2\ ,
\end{eqnarray}
where ${\bf S}=\left({\bm \sigma}_1+{\bm \sigma}_2\right)/2$;
\item{CD terms of NLO ($Q^2$):}
\begin{equation}
v_{\rm NLO}^{\rm CD}= C_0^{\rm IT}\, T_{12} \ ,
\end{equation}
where $T_{12}=3\,\tau_{1z}\tau_{2z}-{\bm \tau}_1\cdot {\bm \tau}_2$ is the isotensor operator;
\item{CD terms of N3LO ($Q^4$):}
\begin{equation}
v_{\rm N3LO}^{\rm CD}({\bf k}, {\bf K})=\left[
C_1^{\rm IT}\,k^2+C_2^{\rm IT}\,k^2\,{\bm \sigma}_1\cdot {\bm \sigma}_2+C_3^{\rm IT}\,S_{12}({\bf k})
+i\,C_4^{\rm IT}{\bf S}\cdot \left({\bf K} \times {\bf k}\right)\right]T_{12} \ .
\end{equation}
\end{itemize}
We note that at N3LO there are four additional CI terms.  Following Ref.~\cite{Piarulli:2016},
we have dropped them, since they lead to operator structures in configuration space which
depend quadratically on the relative momentum operator, and are difficult to implement in quantum
Monte Carlo calculations. Their inclusion was shown to lead to no improvement in the fit to the $2N$ database \cite{Piarulli:2016}. As a matter of fact, three combinations of such terms vanish off the energy shell \cite{Reinert:2017usi} and their effect can be absorbed into a redefinition of the $3N$ interaction \cite{Girlanda:2020}.  We have also ignored five additional charge-symmetry-breaking (CSB)
terms (one at NLO and four at N3LO) in the CD sector.  There is only a single observable
sensitive to these terms, the difference between the $pp$ and $nn$ scattering lengths.
Since the interactions at NLO and N3LO without CSB already give $nn$ scattering lengths
reasonably close to the empirical value (as shown below), we have made no attempt in constraining the
associated LECs, and have therefore set them to zero.
\subsection{Regularization and contact interactions in configuration space}
The contact interactions are regularized by multiplying each term by
a Gaussian cutoff depending only on the momentum transfer $k$ but
which differentiates between the pair isospin $T\,$=$\,0$ and $T\,$=$\,1$ channels, that is
\begin{equation}
\widetilde{C}(k) = {\rm e}^{-  R^2_0 k^2/4}\, P^\tau_0+  {\rm e}^{-  R^2_1 k^2/4}\, P^\tau_1 \longrightarrow
 C(r)=C_0(r)\, P^\tau_0
 +C_1(r)\, P^\tau_1  \ , \qquad C_\alpha(r)= \frac{1}{\pi^{3/2}R_\alpha^3} \, {\rm e}^{-(r/R_\alpha)^2} \ .
 \label{eq:e2.6}
\end{equation}
We have investigated five different combinations of $R_0/R_1$ as listed in Table~\ref{tab:tb1},
and have designated them as models a, b, c, d, and o.  For this latter model (o stands for optimized),
the cutoffs have been determined by constraining them along with the LECs $C_{01}$ and $C_{10}$ in
a LO fit designed to reproduce the $np$ effective range expansions (including
scattering lengths and effective radii) in $S/T\,$=$\,0/1$ and 1/0. We also note that
the relationship between the cutoff $\Lambda_\alpha$ in momentum
space and the cutoff $R_\alpha$ in coordinate space is $\Lambda_\alpha=2/R_\alpha$ (with $\alpha\,$=$\,0$
or 1), and so $\Lambda_0$ and $\Lambda_1$ vary in the ranges (172--263) MeV
and (132--263) MeV as $R_0$ and $R_1$ decrease from 2.3 to 1.5 fm and from 3.0 to
1.5 fm, respectively.
\begin{table}[bth]
\caption{\label{tab:tb1}%
Cutoff values corresponding to models a-d and o.}
\begin{tabular}{c||c|c|c|c|c}
\hline\hline
Model & a & b & c & d & o \\
\hline
$R_0$ (fm) &  1.7 & 1.9 & 2.1 & 2.3&1.54592984  \\
\hline
$R_1$ (fm) &  1.5 & 2.0 & 2.5 & 3.0& 1.83039397 \\
\hline
\hline
\end{tabular}
\end{table}

The coordinate-space representation of the interaction
is written as
\begin{equation}
v=v^{\rm EM}+v^{\rm CI}+v^{\rm CD} \ ,\label{eq:fullpot}
\end{equation}
where $v^{\rm EM}$ is the electromagnetic component, and
\begin{eqnarray}
\label{eq:vr}
v^{\rm CI}&=&v_{\rm LO}^{\rm CI}+v_{\rm NLO}^{\rm CI}+v_{\rm N3LO}^{\rm CI}
=\sum_{l=1}^{11} v^l (r) \, O^l_{12} \ ,\\
v^{\rm CD}&=&v_{\rm NLO}^{\rm CD}+v_{\rm N3LO}^{\rm CD}=\sum_{l=12}^{15} v^l (r) \, O^l_{12}\ .
\end{eqnarray}
The various operator structures of the CI and CD components read
\begin{equation}
\label{eq:op}
O^{l=1,\dots,11}_{12} ={\bf 1}\, ,\, {\bm \tau}_1\cdot {\bm \tau}_2\, , \,
{\bm \sigma}_1\cdot {\bm \sigma}_2\, ,
 {\bm \sigma}_1\cdot {\bm \sigma}_2 \,{\bm \tau}_1\cdot {\bm \tau}_2\, , \,
S_{12}\, , \, S_{12}\, {\bm \tau}_1\cdot {\bm \tau}_2 \, ,\, {\bf L}\cdot {\bf S}\,,\,
{\bf L}\cdot{\bf S}\,{\bm \tau}_1\cdot {\bm \tau}_2\, ,\,
({\bf L}\cdot{\bf S})^2\, ,\, {\bf L}^2\, ,\, 
 {\bf L}^2\, {\bm \sigma}_1\cdot {\bm \sigma}_2 \ ,
 \end{equation}
 and
\begin{equation}
O^{l=12,\dots,15}_{12}=T_{12}
\, ,\, {\bm \sigma}_1\cdot {\bm \sigma}_2\, T_{12}\, , \,S_{12}\, T_{12}\, ,\, {\bf L}\cdot{\bf S}\, T_{12}  \ ,
 \end{equation}
 where $S_{12}$ and ${\bf L}$ denote the tensor and orbital angular momentum
 operators, respectively. Hereafter, we will refer to these operators as
 \begin{equation}
 l=1,\dots, 15 \longrightarrow l=c\,,\, \tau\, , \, \sigma\, ,\, \sigma\tau\, ,\,  t\, ,\, t\tau\, ,\, b\, ,\, b\tau
 \, ,\, bb\, ,\, q\, ,\, q\sigma\, ,\, T,\, \sigma T\, ,\,  tT\, ,\, bT \ .
 \end{equation}
We note that $v^{\rm EM}$ includes the complete electromagnetic interaction
up to terms quadratic in the fine structure constant (first and second
order Coulomb, Darwin-Foldy, vacuum polarization, and magnetic moment
terms), see Ref.~\cite{Wiringa:1994wb} for explicit expressions. The radial functions
$v^l(r)$ multiplying the operators $O^l_{12}$ are given in Ref.~\cite{Piarulli:2014bda} and reported in Appendix~\ref{app:a1} for completeness.
Because of the regularization scheme we have adopted, these functions have an
implicit dependence on the isospin $T$ of the pair.

\section{Fits to the database}
\label{sec:fits}

The (configuration-space) LO, NLO, and N3LO interactions are defined as
\begin{eqnarray}
v_{\rm LO}&=&v^{\rm EM}+v^{\rm CI}_{\rm LO} \ , \\
v_{\rm NLO}&=&v^{\rm EM}+v^{\rm CI}_{\rm LO}+v^{\rm CI}_{\rm NLO}+v^{\rm CD}_{\rm NLO} \ ,\\
v_{\rm N3LO}&=&v^{\rm EM}+v^{\rm CI}_{\rm LO}+v^{\rm CI}_{\rm NLO}+v^{\rm CI}_{\rm N3LO}+v^{\rm CD}_{\rm NLO}+
v^{\rm CD}_{\rm N3LO} \ ,
\end{eqnarray}
where, as already noted, the full EM interaction is retained at each order
(and in all partial waves).  At each order the values of
cutoffs that are considered are those listed in Table~\ref{tab:tb1}.  The LO
interaction involves 2 LECs, the NLO interaction 7 additional LECs in the
CI sector and 1 LEC in the CD sector, and the N3LO interaction further 11
and 4 LECs in the CI and CD sectors, respectively.
As per the operator structure, $v^{\rm CI}_{\rm LO}$
involves the 4 operators $c$, $\tau$, $\sigma$, and
$\sigma\tau$; $v^{\rm CI}_{\rm NLO}$ and $v^{\rm CD}_{\rm NLO}$ involve, respectively, the 7 operators
$c$, $\tau$, $\sigma$, $\sigma\tau$, $t$, $t\tau$, and
$b$, and the single operator $T$; $v^{\rm CI}_{\rm N3LO}$ and $v^{\rm CD}_{\rm N3LO}$ involve, respectively, the 11 operators
$c$, $\tau$, $\sigma$, $\sigma\tau$, $t$, $t\tau$,
$b$, $b\tau$, $bb$, $q$, and $q\sigma$, and the 4 operators $T$, $\sigma T$, $tT$\, $bT$. However, because of the isospin dependence of the radial functions $v^l(r)$, the
interactions
$v_{\rm NLO}^{\rm CI}$ and $v_{\rm N3LO}^{\rm CI}$ effectively also include, respectively, the $b\tau$ operator, and the ${\bm \tau}_1\cdot{\bm \tau}_2$-dependent
$bb$, $q$, and $q\sigma$ operators.
Lastly, the values
adopted for the proton and neutron masses are, respectively, $938.27192$ MeV and
$939.56524$ MeV, and $\hbar c$ is taken as 197.32697 MeV$\,$fm.

The 2 LECs in the LO interactions are determined by reproducing the singlet $np$
scattering length ($^1a_{np}$) and the deuteron binding energy ($B_d$) in models
a-d.  In model o, the cutoff radii along with the LO LECs have been
constrained by fitting the $np$ scattering lengths and effective radii in the singlet and triplet
channel, and the deuteron binding energy.  Their values are listed in Table~\ref{tab:tb2}.  
The NLO and N3LO interactions are fitted to $np$ and $pp$
scattering data (including normalizations), as assembled in the Granada
database~\cite{Navarro:2013,Navarro:2014,Navarro:2014b}, over
the laboratory energy range 0--15~MeV at NLO and 0--25~MeV at N3LO, and, simultaneously, to $B_d$.
The corresponding LECs are reported in Table~\ref{tab:tb2anew} at NLO and Table~\ref{tab:tb2b} at N3LO,
in Appendix~\ref{app:a2}.  The optimization of the objective function $\chi^2$ with respect to the
LECs is carried out with the Practical Optimization Using No Derivatives (for Squares),
POUNDERS~\cite{Kortelainen:2010}.
\begin{table}[t]
\caption{\label{tab:tb2}%
The LO LECs determined by reproducing the $np$ singlet scattering length and deuteron binding energy
as obtained for models a-d; for model o, the cutoff radii along with the LO LECs have been
constrained by fitting the $np$ scattering lengths and effective radii in the singlet and triplet
channel, and the deuteron binding energy.}
\begin{ruledtabular}
\begin{tabular}{lccccc}
Model & a& b & c & d & o \\
\hline\hline
$C_{01}$(fm$^2$) & --.438524414E+01& --.572220536E+01 & --.700250932E+01 & --.822926713E+01 & --.527518671E+01  \\
$C_{10}$(fm$^2$) & --.800783936E+01& --.934392090E+01  & --.107734100E+02  & --.122993164E+02  &--.704040080E+01   \\
%
\end{tabular}
\end{ruledtabular}
\end{table}
We list the numbers of $np$, $pp$, $np+pp$ data (including normalizations)
and corresponding $\chi^2$/datum for all models in Table~\ref{table:chi2}.
The NLO and N3LO fits are optimized by minimizing the $\chi^2$
corresponding to the total number of $np\,$+$\,pp$ data.  The numbers of
data points change slightly for each of the various models because of
fluctuations in the number of normalizations, see Ref.~\cite{Piarulli:2014bda}
for more details on the fitting procedure.  Finally, in Table~\ref{table:chi2}
we also report the $\chi^2$/datum to the $np$ data in the laboratory
energy range 0--1 MeV for the LO models.  We stress that these $\chi^2$
values do not result from fits, but rather correspond to
the sets of LECs as determined in Table~\ref{tab:tb2}.  We do
not report the $\chi^2$/datum values relative to the $pp$ data, since they
are in the thousands to tens of thousands (the number
of $pp$ data points in 0--1 MeV is about 160), and therefore meaningless. 
\begin{table*}[h]
\caption{Values of the $\chi^2$/datum at LO, NLO and N3LO.
The $\chi^2$/datum values reported at LO over the lab-energy range \hbox{$T_{\rm lab}\,$=$\,$0--1}
MeV are obtained with the LECs of Table~\ref{tab:tb2}.  The NLO (N3LO) fits are performed
over the range \hbox{$T_{\rm lab}\,$=$\,$0--15} (0--25) MeV; $N_{np}$, $N_{pp}$, and $N$
denote, respectively, the total number of $np$, $pp$, and $np\,$+$\,pp$ data, including
observables and normalizations.  The NLO and N3LO fits are carried out by
enforcing that the deuteron binding energy be reproduced exactly, and 
are optimized by minimizing the $\chi^2$ corresponding to the total number of $np\,$+$\,pp$ data.
}
\begin{center}
\begin{ruledtabular}
\begin{tabular}{ccccccccc}
Model&  order&  $T_{\rm lab}$ (MeV) & $N_{np}$ & $\chi^2(np)$/datum &
$N_{pp}$ & $\chi^2(pp)$/datum& $N$ & $\chi^2$/datum\\
\hline
a & LO   &  0--1 &  91  & 5.54   &   157&      &  248  &     \\
            & NLO   &  0--15 &  381&   1.83 & 394  &  1.53   &  776 & 1.67   \\
            & N3LO &   0--25  & 643  & 1.60  &  451 & 1.24 &  1096 & 1.45\\
\hline
b & LO   &  0--1 & 91   & 37.6   &  157 &      &   248 &     \\
            & NLO   &  0--15 &   382& 1.39   & 395 & 1.09     & 778   & 1.24   \\
            & N3LO &   0--25  & 646    & 1.42   & 452  &  1.06 & 1099  & 1.27  \\
\hline
c & LO   &  0--1 &91   & 24.8    & 157  &      &  248  &     \\
            & NLO   &  0--15 &  378 & 2.34   &   392 &    1.97  & 771  & 2.15  \\
            & N3LO &   0--25  & 645   & 1.83    & 453  & 1.33 &  1099 &  1.62 \\
\hline
d & LO   &  0--1 &  91 & 41.2   &   157 &      &  248  &     \\
            & NLO   &  0--15 & 377 & 10.2   &  392 &    6.88  & 770  & 8.51   \\
            & N3LO &   0--25  & 638   & 2.03  &  446 & 8.09  & 1085  & 4.52 \\
            \hline
o& LO   &  0--1 &  91 &  2.16   &   157 &      &  248  &     \\
            & NLO   &  0--15 & 382 & 1.27   &  394 &    1.08  & 777  & 1.17   \\
            & N3LO &   0--25  &  650  &  1.25 &  452 & 1.10  & 1103 & 1.19\\
\end{tabular}
\end{ruledtabular}
\label{table:chi2}
\end{center}
\end{table*}

The $\chi^2$ improves slightly or remains essentially unchanged in going from NLO to
N3LO, albeit the number of data points included in the fits increases from about
780 at 15~MeV to about 1,100 at 25~MeV; the $\chi^2$ improvement is drastic,
by about a factor of 2 for model d, corresponding to
$R_0/R_1\,$=$\,$2.3/3.0 fm.  But for this model, all $\chi^2$ at N3LO are
well below 2.  Even in the limited range of laboratory energy we have considered,
the data points number in the several hundreds, and consist primarily
of differential cross sections.  The $\chi^2$ values at NLO and N3LO 
relative to the $pp$ data are generally significantly better than those relative to $np$
data, except again at N3LO for model d for which this trend is reversed (it worthwhile
reiterating here that the fits are optimized by minimizing the $\chi^2$
relative to the $np$ {\it and} $pp$ data).

We conclude this section by noting that in an early exploratory phase of the present work,
we considered interactions regularized by a single cutoff function, namely without
differentiating between pairs in isospin $T\,$=$\,$0 and 1.  This is equivalent to setting
$R_0\,$=$\,R_1\,$=$\,R$, and
\begin{equation}
 C(r)=\frac{1}{\pi^{3/2}R^3} \, {\rm e}^{-(r/R)^2} \ .
\end{equation}
Both NLO and N3LO interactions were fitted to the database over the
energy range 0--15 MeV (albeit the deuteron binding energy was not included
in the fits), and with cutoff $R$ varying between
1.0 and 2.5 fm, see Table~\ref{table:chi2old}.  We found the $\chi^2$
corresponding to the $np$ data fits to be rather large when the cutoff
$R$ was taken either too small $R\lesssim 1.0$ fm or too large $R\gtrsim 2.5$ fm.
Moreover, the deuteron binding energy was generally poorly reproduced at
both NLO and N3LO; for example, with $R\,$=$\,2.5$ fm it was calculated to be
1.243 (1.312) MeV at NLO (N3LO).  This led us to (i) introduce two cutoffs
differentiating between $T\,$=$\,$0 and 1 pairs in order to allow for
different ranges in these channel interactions, (ii) restrict the variability of
the $R_0$ cutoff between 1.5 fm and 2.3 fm, in order to improve the $\chi^2$,
and (iii) include in the fits the deuteron binding energy which, because
of the small experimental error associated with it, puts a very tight constraint 
on the $\chi^2$.
\begin{table*}[t]
\caption{Values of the $\chi^2$/datum at NLO and N3LO obtained by setting
$R_0\,$=$\,R_1\,$=$\,R$, namely without differentiating the range of the
interactions between $T\,$=$\,$0 and $T\,$=$\,$1 pairs.  Remaining notation is
as in Table~\ref{table:chi2}. The NLO and N3LO fits are optimized
by minimizing the $\chi^2$ corresponding to the total number of $np+pp$ data
over the same range \hbox{$T_{\rm lab}\,$=$\,$0--15} MeV.  Note that the deuteron
binding energy was not included in the fits.
}
\begin{center}
\begin{ruledtabular}
\begin{tabular}{ccccccccc}
$R$ (fm) &  order&  $T_{\rm lab}$ (MeV) & $N_{np}$ & $\chi^2(np)$/datum &
$N_{pp}$ & $\chi^2(pp)$/datum& $N$ & $\chi^2$/datum\\
\hline
1.0 & NLO   &  0--15 & 375 & 15.5  & 390 &    8.61 & 776  & 12.0   \\
      & N3LO &   0--15  & 366  & 5.95 & 392 & 3.96  & 758& 4.92 \\
\hline
1.5 & NLO   &  0--15 & 366 &3.32  & 392 &  1.49  & 758  &  2.38 \\
      & N3LO & 0--15   & 369   & 1.44  & 395  & 1.64  & 764 & 1.55 \\
\hline
2.0 & NLO   & 0--15   & 367 & 2.78 &  391 & 1.46 & 758 &  2.10 \\
      & N3LO & 0--15   & 367  & 1.66 & 393 & 0.95 & 760 & 1.29 \\
\hline
2.5 & NLO   & 0--15   &  373 & 9.75  &393  & 2.19    & 766 & 5.87 \\
      & N3LO & 0--15   & 374  & 3.48 & 392 & 1.85  & 766 & 2.64 \\
\end{tabular}
\end{ruledtabular}
\label{table:chi2old}
\end{center}
\end{table*}
\subsection{Deuteron properties, effective range parameters, and phase shifts}
Deuteron properties obtained at NLO and N3LO are reported in Table~\ref{tab:tb5new}
and compared to available experimental values.  The binding energy $B_d$ is fitted exactly
and includes the contributions (about 20 keV) of electromagnetic interactions, among which
the largest is that due to the magnetic moment term.  The asymptotic D/S ratio $\eta$ is reasonably close to experiment for models a, b, and d, but is significantly
overpredicted and underpredicted in model c and o, respectively.
\begingroup
\squeezetable
\begin{table}[h]
\caption{\label{tab:tb5new}
Deuteron binding energy $B_d$ (in MeV), D-to-S state ratio $\eta$, and D-state
probability ($P_D$) obtained at NLO and N3LO; the experimental values~\cite{Ericson:1983,Rodning:1990,Huber:1998,Martorell:1995}
are, respectively, $B_d\,$=$\,2.224575(9)$ MeV and $\eta\,$=$\,$$0.0256(4)$.  The superscript $^*$ indicates that the
corresponding observable is fitted.}
\begin{tabular}{l||S|S||S|S||S|S||S|S||S|S}
Model
    & \multicolumn{2}{l} {\hspace{1.5cm}a} & \multicolumn{2}{l}{\hspace{1.5cm}b} & \multicolumn{2}{l}{\hspace{1.5cm}c}
    & \multicolumn{2}{l} {\hspace{1.4cm} d} & \multicolumn{2}{l} {\hspace{1.4cm} o} \\
    \hline
          & NLO  & \hbox{N3LO}  & NLO  & \hbox{N3LO} & NLO  & \hbox{N3LO}&NLO & \hbox{N3LO}&NLO & \hbox{N3LO}\\
       \hline
$B_{d}$ (MeV) & 2.2246$^*$ & 2.2246$^*$  &  2.2246$^*$  &  2.2246$^*$ & 2.2246$^*$& 2.2246$^*$  & 2.2246$^*$  &     2.2246$^*$&2.2246$^*$&2.2245$^*$	\\
\hline
$\eta$               &   0.0233 &  0.0235&0.0237& 0.0238&   0.0373 & 0.0351  &  0.0231 & 0.0226 & 0.0169&0.0170\\
\hline
$P_D$ (\%)               &   2.93 &  2.96 & 2.24    & 2.30& 4.11   &  4.39  &   1.20  &   1.84 & 2.04&1.93\\
\hline
\end{tabular}
\end{table}
\endgroup
\begin{figure}[bth]
\includegraphics[width=2.4in]{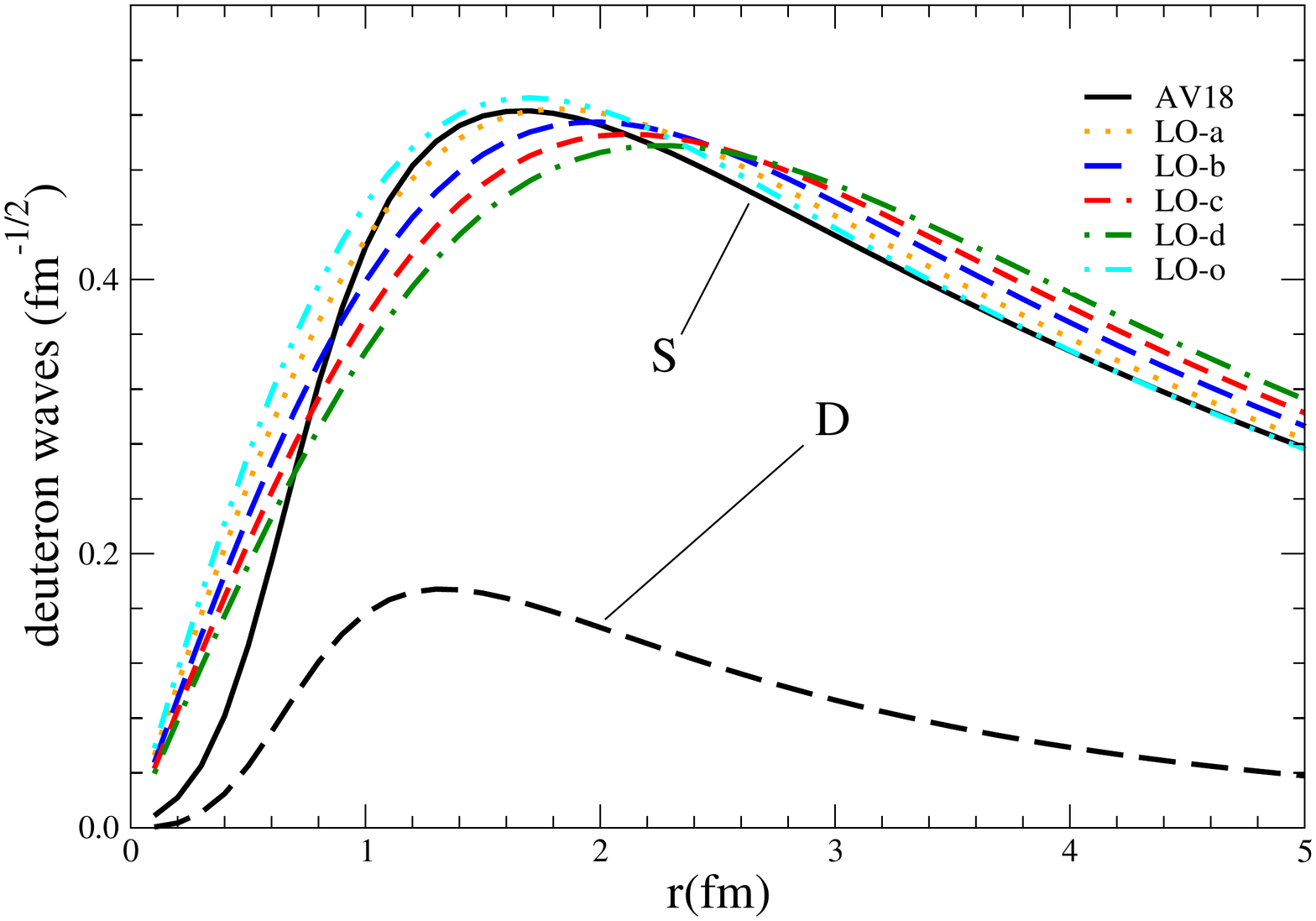}\includegraphics[width=2.4in]{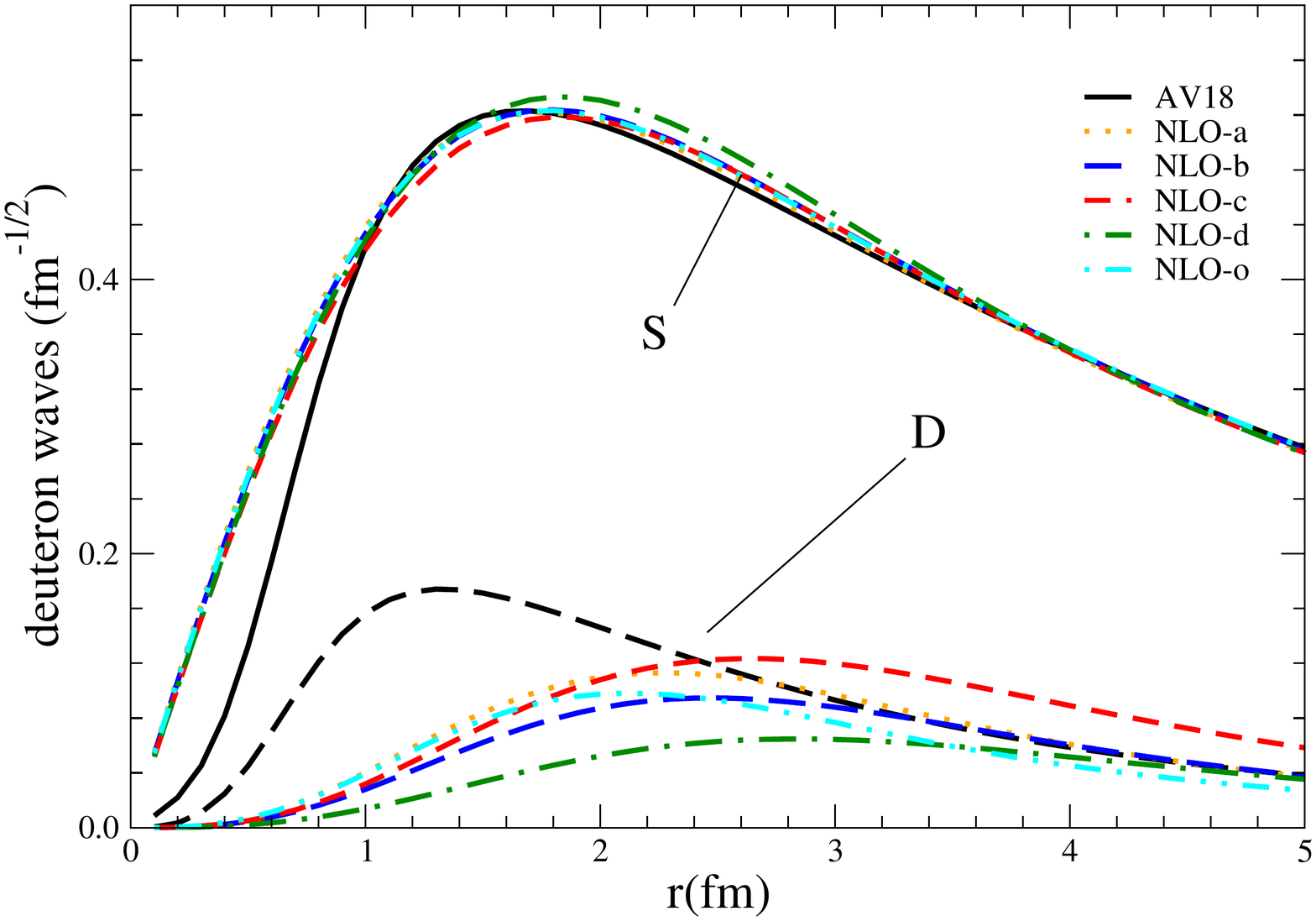}\includegraphics[width=2.4in]{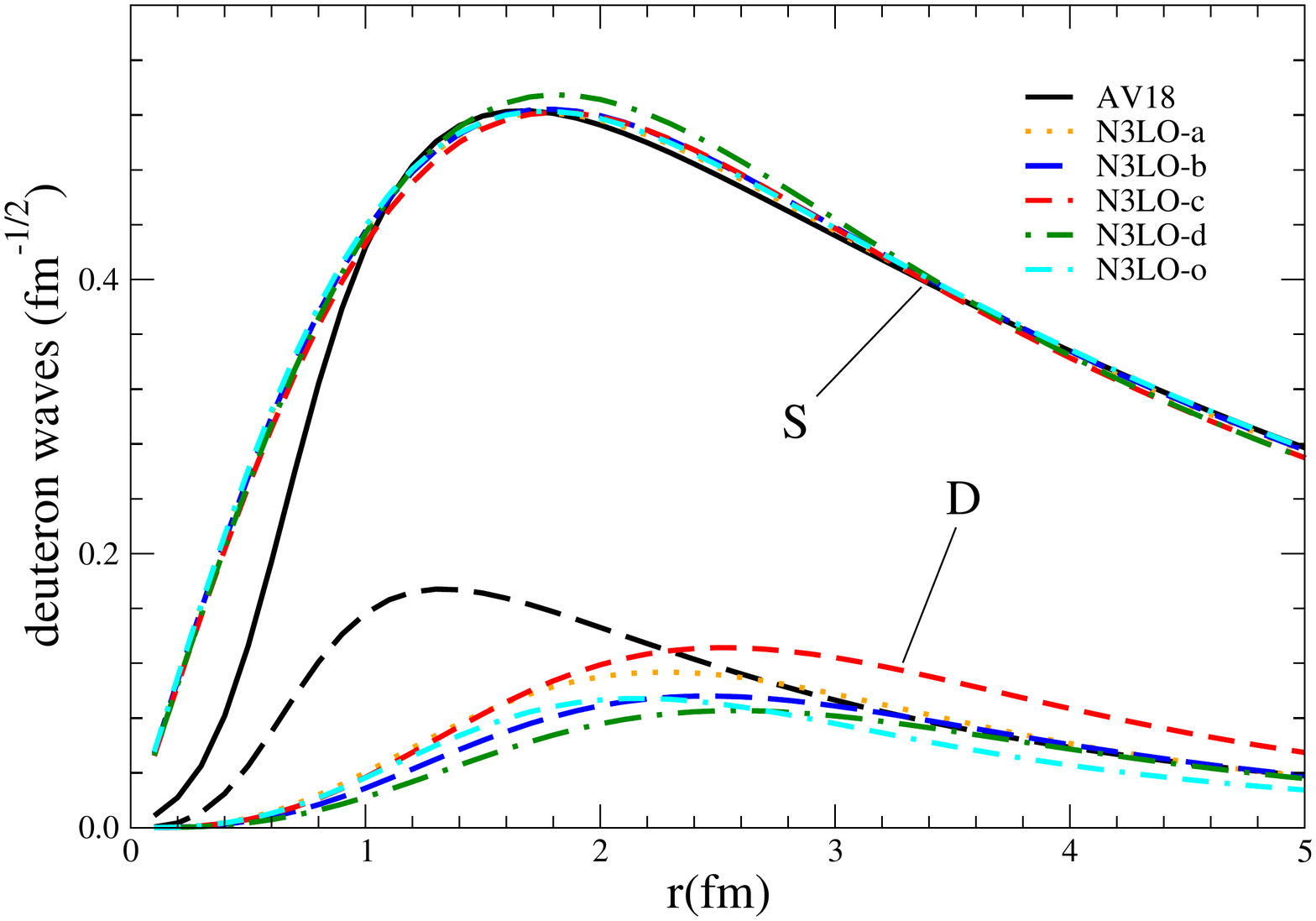}
\caption{(Color online). The deuteron S-wave radial functions at LO (left panel),
corresponding to the LECs of Table~\ref{tab:tb2},
and deuteron S- and D-wave radial functions at NLO (middle panel) and N3LO (right panel),
corresponding to the best fits of Table~\ref{table:chi2}, are compared to those of the AV18.
Note that at LO the tensor term from the $np$ magnetic-moment interaction
induces tiny D-wave components, which are not shown.}
\label{fig:deut_lo}
\end{figure}
Deuteron waves at LO, NLO, and N3LO are shown in Fig.~\ref{fig:deut_lo}.  They are
compared to the S- and D-wave obtained with the AV18~\cite{Wiringa:1994wb} for reference.
Note that the tensor term in the $n$-$p$ magnetic-moment interaction
induces at LO tiny D-waves, which are not displayed in Fig.~\ref{fig:deut_lo}.
The NLO and N3LO D-waves in all models are smaller than the AV18
D-wave, and are pushed out relative to it.  By contrast, the LO, NLO, and N3LO S-waves
at short range are significantly larger than the AV18 S-wave, reflecting
the absence of a repulsive core in the contact interactions.
\begin{figure}[bth]
\includegraphics[width=5.75in]{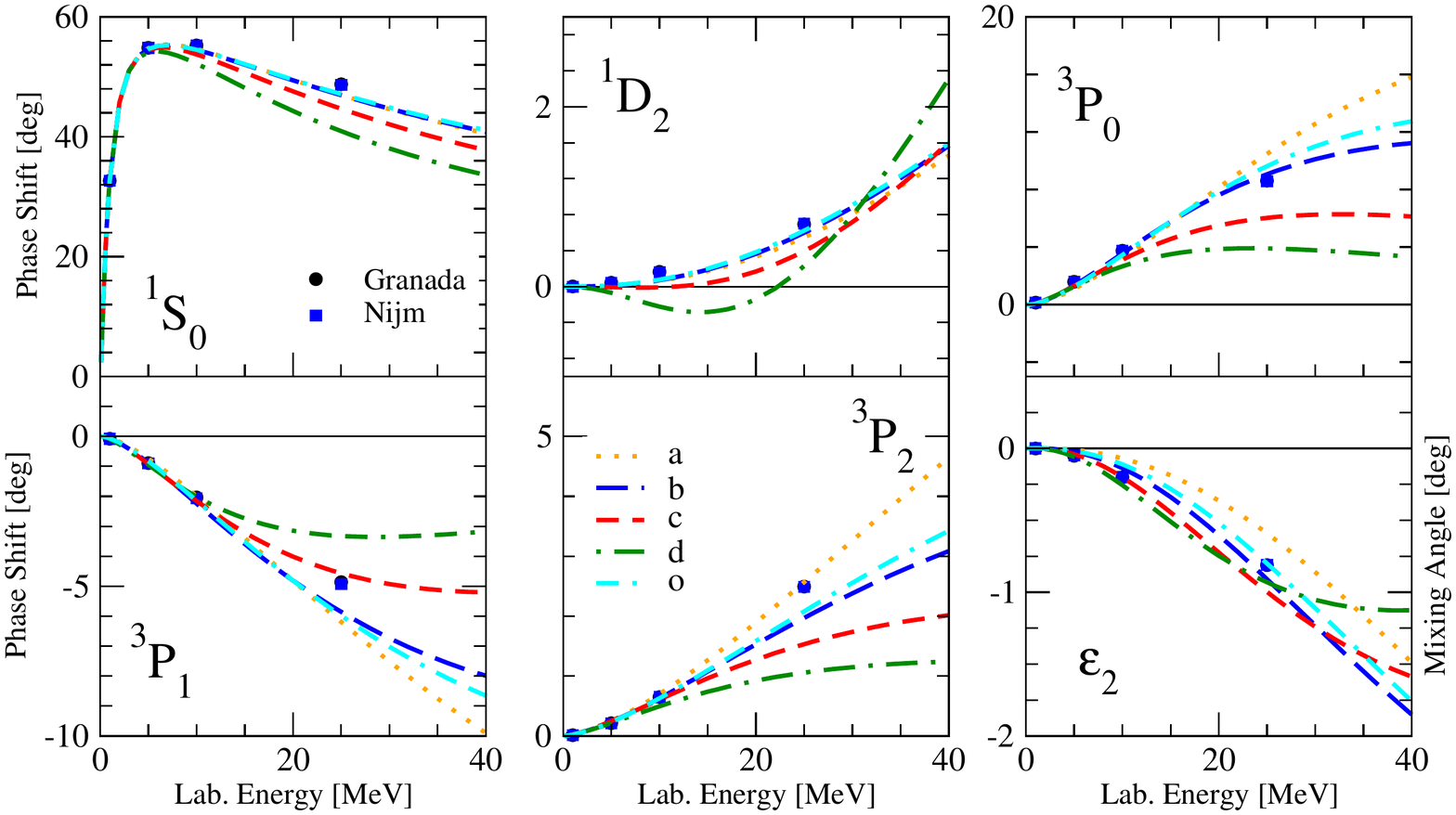}\\
\includegraphics[width=5.75in]{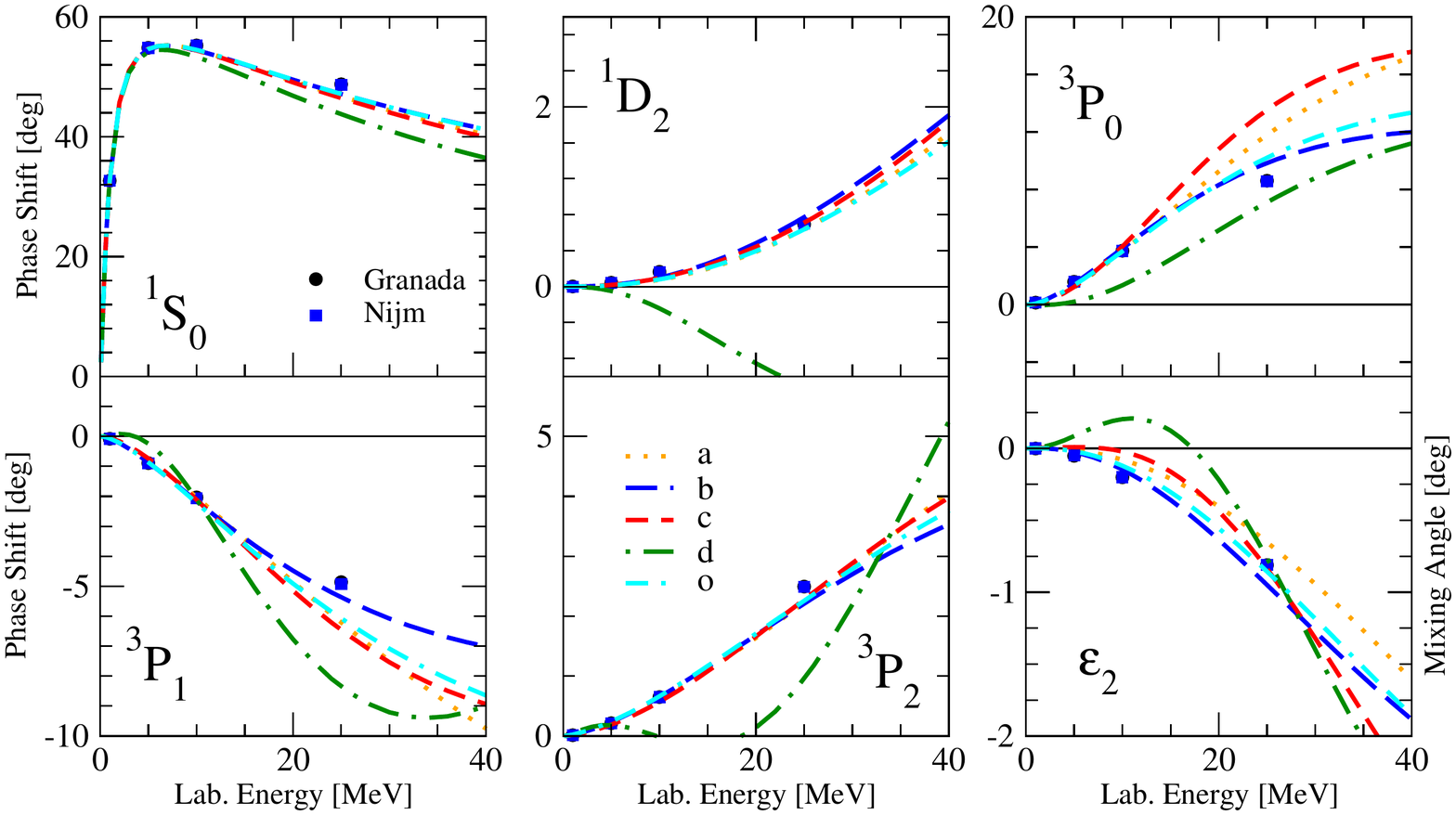}
\caption{(Color online). Phase shifts in $pp$ channels at NLO (top panel) and N3LO (bottom panel)
corresponding to the best fits of Table~\ref{table:chi2} are compared to the results of the Nijmegen
and Granada partial-wave analyses.}
\label{fig:pp_nlo}
\end{figure}
\begin{figure}[bth]
\includegraphics[width=5.75in]{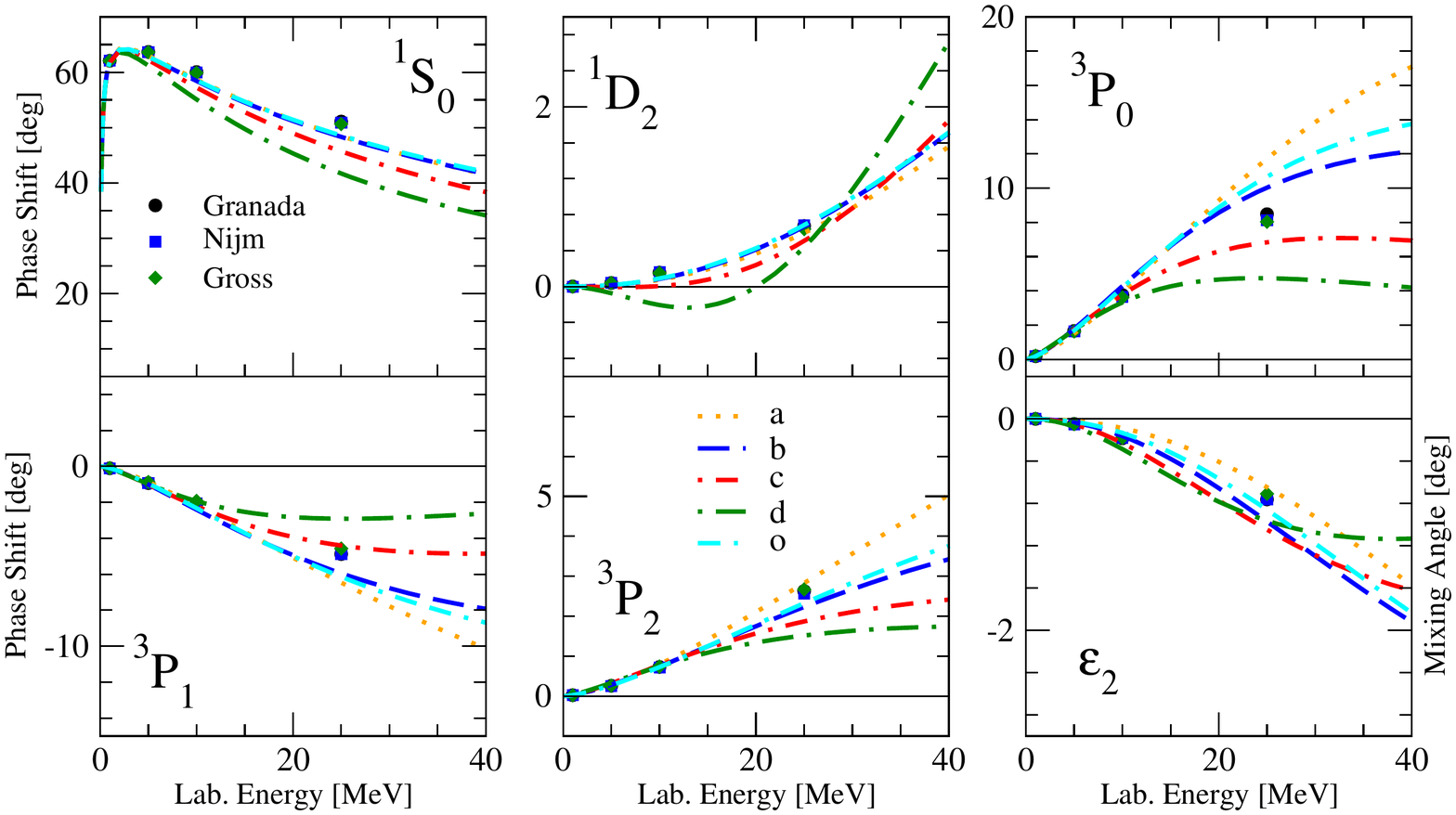}\\
\includegraphics[width=5.75in]{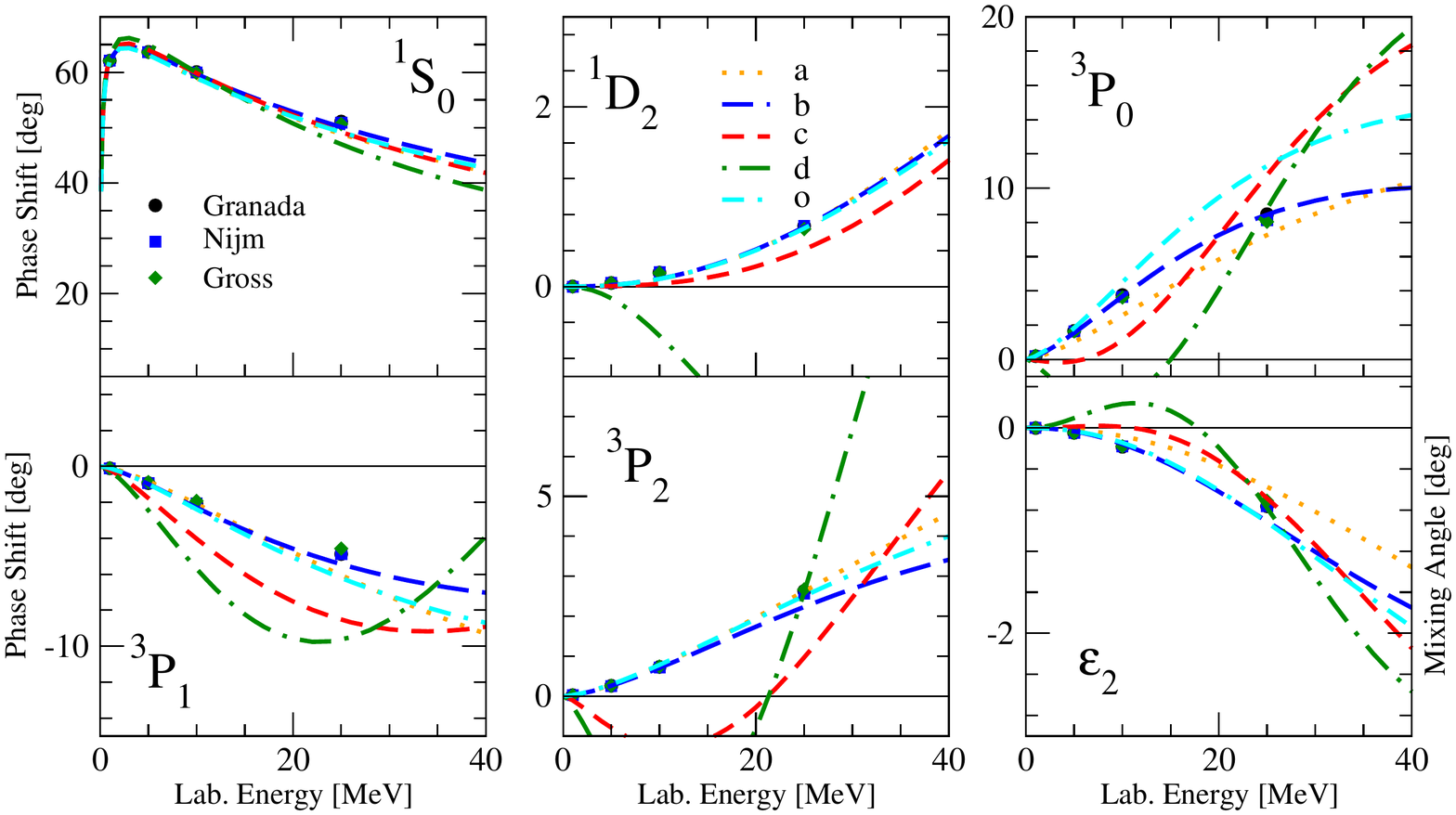}
\caption{(Color online). Phase shifts in isovector $np$ channels at NLO (top panel) and N3LO (bottom panel)
corresponding to the best fits of Table~\ref{table:chi2} are compared to the results of the Nijmegen,
Granada, and Gross partial-wave analyses.}
\label{fig:np_iv_nlo}
\end{figure}
\begin{figure}[bth]
\includegraphics[width=5.75in]{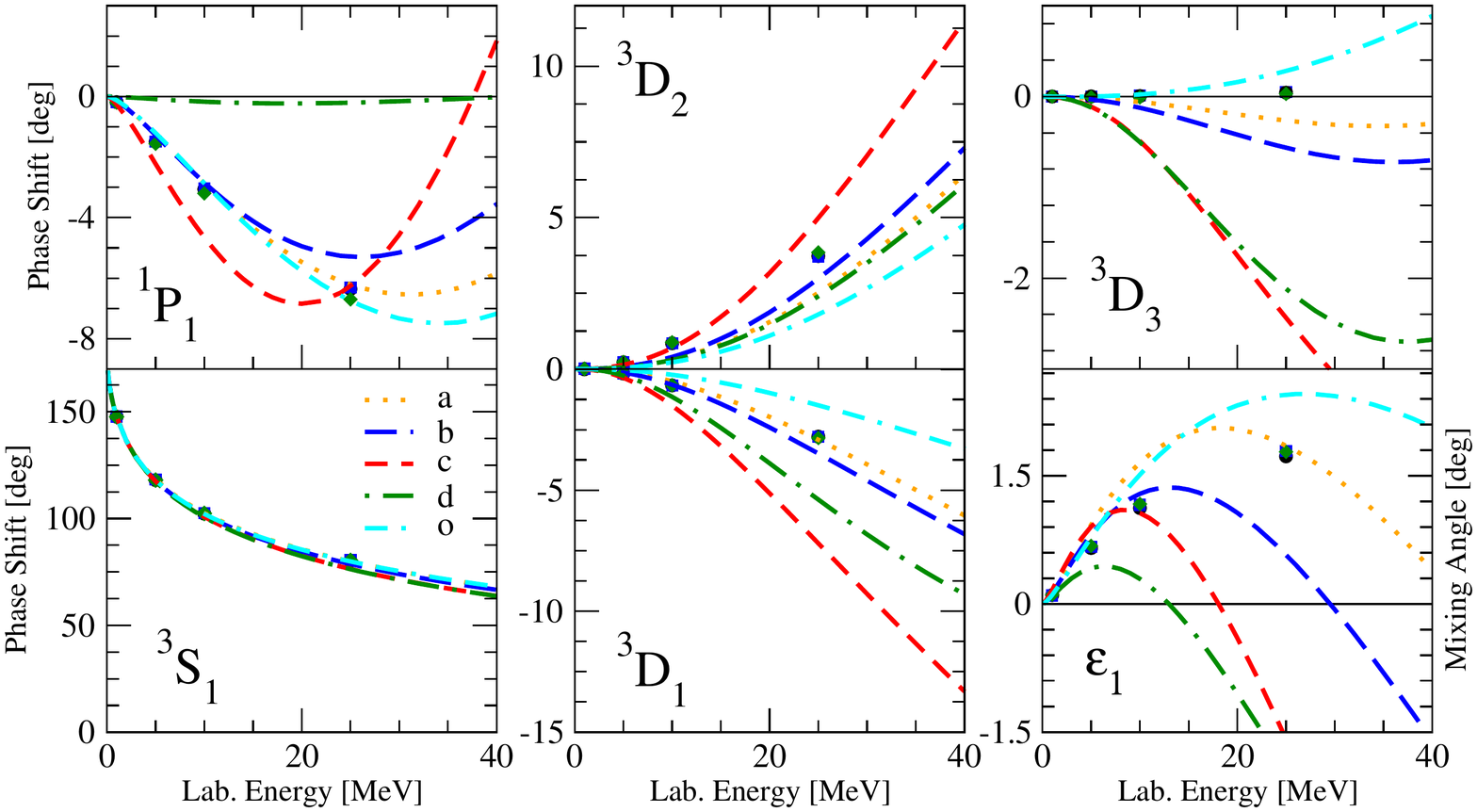}\\
\includegraphics[width=5.75in]{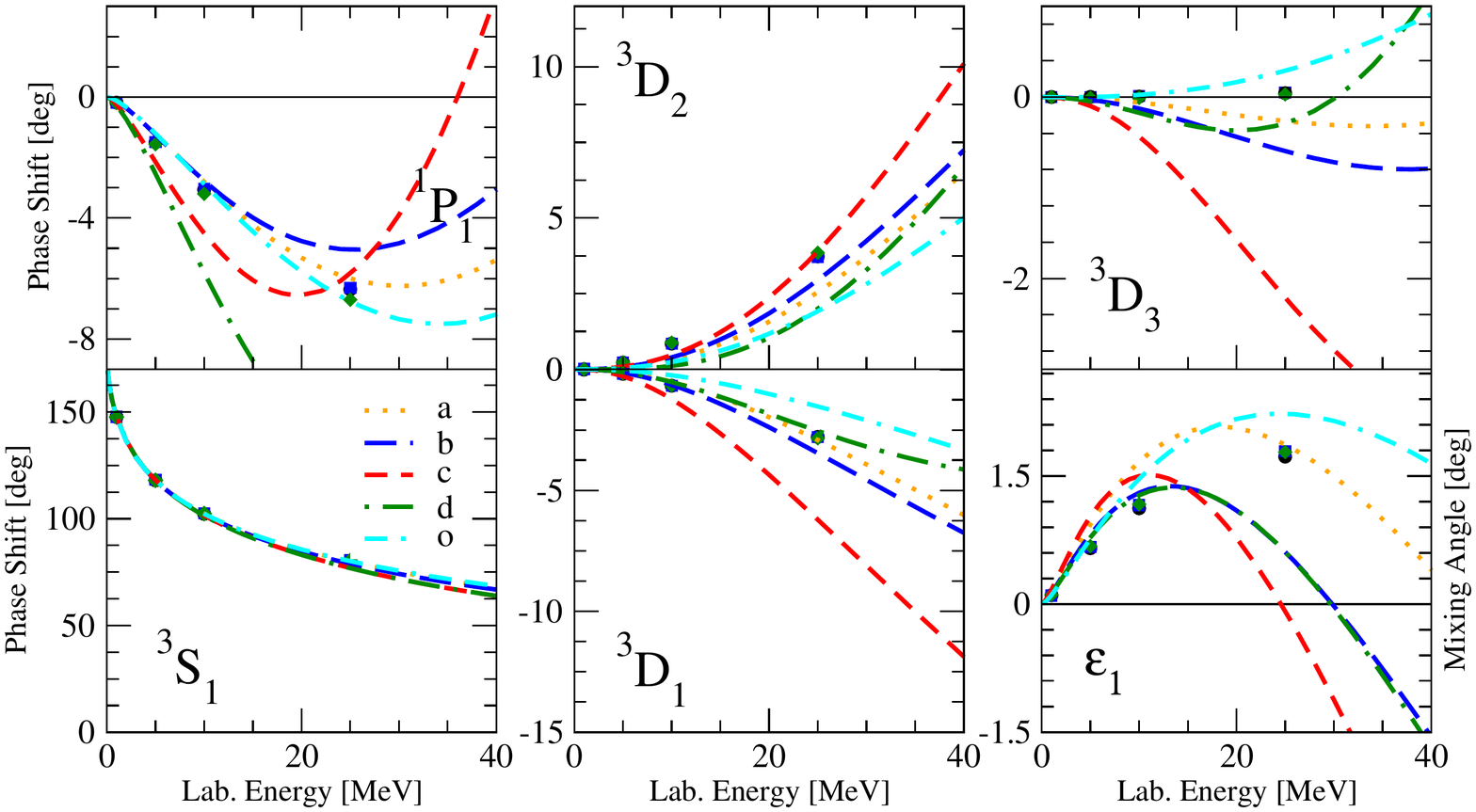}
\caption{(Color online).  Phase shifts in isoscalar $np$ channels at NLO (top panel) and
N3LO (bottom panel) corresponding to the best fits of Table~\ref{table:chi2} are compared to
the results of the Nijmegen, Granada, and Gross partial-wave analyses. }
\label{fig:np_is_nlo}
\end{figure}

The effective range parameters at LO, NLO, and N3LO are given in
Tables~\ref{tab:tb3lo},~~\ref{tab:tb3nlo}, and~\ref{tab:tb3new}, respectively,
where they are compared to experimental values.  The experimental
value of the singlet $np$ scattering length is reproduced exactly by design in the
case of the LO models a-d.  However, the LO model o has been constrained
to reproduce the $np$ effective range expansions in the singlet and triplet
channels as well as the deuteron binding energy. 
At all orders of the power counting, the singlet and triplet
$np$, and singlet $pp$ and $nn$, scattering lengths are calculated with the inclusion
of electromagnetic interactions.  Without the latter, the effective-range function is
simply given by $F(k^2)=k\, \cot\, \delta=-1/a+r\, k^2/2$ up to terms linear in $k^2$.
In the presence of electromagnetic interactions, a more complicated effective-range
function must be used; it is given explicitly in Appendix D of Ref.~\cite{Piarulli:2014bda},
along with relevant references. 

The predicted scattering lengths at NLO and N3LO are typically within a \%
of the experimental values for all models considered.  However, the effective
radii display more variability from model to model, but are all reasonably close
to experiment.
\begin{table*}[bth]
\caption{
The singlet and triplet $np$, and singlet $pp$ and $nn$, scattering lengths and
effective radii obtained at LO;
experimental values are from Refs.~\cite{Bergervoet:1988,Sanden:1983,Chen:2008,Miller:1990,Machleidt:2000ge}.
The superscript $^*$ indicates the
corresponding observable is fitted.}
\label{tab:tb3lo}
\begin{ruledtabular}
\begin{tabular}{ldddddd}
\textrm{}&
\multicolumn{1}{r}{\textrm{Experiment}}&
\multicolumn{1}{c}{\textrm{a}}&
\multicolumn{1}{c}{\textrm{b}}&
\multicolumn{1}{c}{\textrm{c}}&
\multicolumn{1}{c}{\textrm{d}}&
\multicolumn{1}{c}{\textrm{o}}\\
\colrule
$^{1}a_{pp}$ 	&-7.8063(26)	  &-8.1234  &  -8.8643	&-9.5462  & -10.1886 & -8.6207 \\
		        &-7.8016(29)	  &   &	&	     &    &  \\
$^{1}r_{pp}$	& 2.794(14)   	  &  2.180 &	2.909& 3.640& 4.371	&2.662 \\
 			& 2.773(14)   	  &   &	& 			& & \\
$^{1}a_{nn}$ 	&-18.90(40)	  & -22.13 &-22.68	& -23.01	&-23.21 & -22.53 \\
$^{1}r_{nn}$	& 2.75(11)  	  & 2.26  &3.05	& 3.87    &		4.71& 2.78\\
$^{1}a_{np}$	&-23.740(20) 	  & -23.740^* &	-23.740^*&	-23.740^*	&-23.740^*& -23.740^*\\
$^{1}r_{np}$	& 2.77(5)		  &  2.25 &	3.04& 	2.65	&4.69	& 2.77^*\\
$^{3}a_{np}$	& 5.419(7)  	  &  5.515 & 5.650	& 	5.783& 5.913& 5.410^*\\
$^{3}r_{np}$	& 1.753(8)   	  & 1.89 & 2.06	& 	2.21& 2.36&1.757^*	\\
\end{tabular}
\end{ruledtabular}
\end{table*}
\begin{table*}[bth]
\caption{Same as in Table~\ref{tab:tb3lo} but at NLO.}
\label{tab:tb3nlo}
\begin{ruledtabular}
\begin{tabular}{ldddddd}
\textrm{}&
\multicolumn{1}{r}{\textrm{Experiment}}&
\multicolumn{1}{c}{\textrm{a}}&
\multicolumn{1}{c}{\textrm{b}}&
\multicolumn{1}{c}{\textrm{c}}&
\multicolumn{1}{c}{\textrm{d}}&
\multicolumn{1}{c}{\textrm{o}}\\
\colrule
$^{1}a_{pp}$ 	&-7.8063(26)	 & -7.7489  &  -7.7557	& -7.7463 &   -7.7119  &  -7.7570\\
		        &-7.8016(29)     &   &	&	   & \\
$^{1}r_{pp}$	& 2.794(14)   	  & 2.649 & 2.676	& 2.622		& 2.439	& 2.682	\\
 			& 2.773(14)   	  &   &	& 		&	\\
$^{1}a_{nn}$ 	&-18.90(40)	  &  -17.19 &-17.21	&-16.90		&-16.45&-17.23 	\\
$^{1}r_{nn}$	& 2.75(11)  	  &   2.78 & 2.80	&     	2.79 &	2.72& 2.80\\
$^{1}a_{np}$	&-23.740(20) 	  & -23.765 &	-23.740 &-23.746	&-23.740& -23.738	\\
$^{1}r_{np}$	& 2.77(5)		  &  2.71 &2.75	& 2.78	&2.78& 	2.75\\
$^{3}a_{np}$	& 5.419(7)  	  &   5.392 &5.424 & 5.418	& 	5.415&5.426 \\
$^{3}r_{np}$	& 1.753(8)   	  &  1.746 &1.796 & 1.831	&  1.838 &1.782 \\
\end{tabular}
\end{ruledtabular}
\end{table*}
\begin{table*}[bth]
\caption{Same as in Table~\ref{tab:tb3lo} but at N3LO.}
\label{tab:tb3new}
\begin{ruledtabular}
\begin{tabular}{ldddddd}
\textrm{}&
\multicolumn{1}{r}{\textrm{Experiment}}&
\multicolumn{1}{c}{\textrm{a}}&
\multicolumn{1}{c}{\textrm{b}}&
\multicolumn{1}{c}{\textrm{c}}&
\multicolumn{1}{c}{\textrm{d}}&
\multicolumn{1}{c}{\textrm{o}}\\
\colrule
$^{1}a_{pp}$ 	&-7.8063(26)	  & -7.7539    &  -7.7634 	&-7.7554 	  &-7.7730   &-7.7590   \\
		        &-7.8016(29)	  &   &	&	     &\\
$^{1}r_{pp}$	& 2.794(14)   	  &   2.669 &2.709 & 2.674 &	2.744 &	 2.690\\
 			& 2.773(14)   	  &   &	& 			&      &     \\
$^{1}a_{nn}$ 	&-18.90(40)	  & -17.15&-17.22 	& -17.13	&-16.72& -17.23\\
$^{1}r_{nn}$	& 2.75(11)  	  & 2.80  & 2.83	& 2.80   &		2.90& 2.81       \\
$^{1}a_{np}$	&-23.740(20) 	  & -23.760  &	-23.745&	-23.780	&-23.794&  -23.739    \\
$^{1}r_{np}$	& 2.77(5)		  &  2.68 &	2.60& 	2.49  &2.15&      2.70   	\\
$^{3}a_{np}$	& 5.419(7)  	  &  5.397 & 5.415& 	5.366  & 5.363 &   5.422   \\
$^{3}r_{np}$	& 1.753(8)   	  & 1.754  & 1.784	&1.769& 1.776&    1.775  	\\
\end{tabular}
\end{ruledtabular}
\end{table*}

The $pp$ and (isovector and isoscalar) $np$ S-, P-, and D-wave phase shifts obtained
with the NLO and N3LO interaction models up to a laboratory energy of 40 MeV,
are displayed in Figs.~\ref{fig:pp_nlo}--\ref{fig:np_is_nlo}, and are compared to
partial-wave analyses (PWAs) by the Nijmegen~\cite{Stoks:1993},
Granada~\cite{Navarro:2013,Navarro:2014,Navarro:2014b}, and
Gross-Stadler~\cite{Gross:2008} groups.  The Gross-Stadler PWA is limited
to $np$ data only.  The $pp$ phases are relative to electromagnetic
functions~\cite{Piarulli:2014bda}, while the $np$ ones are relative
to spherical Bessel functions.  Except for the S-wave phase shifts, there is a rather large
spread in the P- and D-wave phase shifts and mixing angles among the different
models.  This spread does not appear to be reduced in going from NLO to N3LO,
although it is worthwhile reiterating here that the fits to the database were restricted
to a rather low upper limit in the energy range and that in such a range the
data, which consist primarily of differential cross sections, are not very sensitive
to higher-order partial waves.

\section{Binding energies of light and medium-weight nuclei}
\label{sec:a34}

\begin{table*}[bth]
\caption{Values obtained at LO, NLO, and N3LO for the LEC $c_E$ in the $3N$ contact interaction,
corresponding to cutoffs $R_3\,$=$\, 1.0$, 1.5, 2.0, and 2.5 fm.  Each combination is constrained to reproduce
the $^3$H binding energy in HH calculations. 
Parts (A) and (B) report the $c_E$ values obtained by either ignoring (A)
or retaining (B) the full $v^{EM}$ in the $^3$H calculations.}
\begin{center}
\begin{ruledtabular}
\begin{tabular}{cc dddd}
\multicolumn{6}{c}{(A)}\\
Model&  order&  1.0\, {\rm fm}  & 1.5 \, {\rm fm} & 2.0\, {\rm fm} &  2.5\, {\rm fm}\\
\hline\hline
a & LO   & 1.8354  & 4.6301  & 11.6871 & 27.4702 \\
  & NLO  & 0.14877 & 0.38897 & 0.97039 & 2.24176 \\
  & N3LO & 0.14478 & 0.37956 & 0.94411 & 2.18030  \\
  \hline
b & LO  &  0.02828 & 0.06903  & 0.16387  & 0.36545 \\
  & NLO  & 0.33198 & 0.86155 & 2.14635 &  4.95746 \\
  & N3LO & 0.47281 & 1.23309 & 3.09130 &  7.17598 \\
  \hline
c & LO   & -2.09231 & -5.37280 & -12.4415 & -26.8473 \\
  & NLO  & -0.47519 & -1.23710 & -3.02891 & -6.87885 \\
  & N3LO &  0.01615 &  0.04168 &  0.10266 & 0.23501 \\
  \hline
d & LO   & -3.89132 & -10.9436 & -25.3577 & -53.7786 \\
  & NLO  & -0.58694 & -1.46947 & -3.50072 & -7.80518 \\
  & N3LO &  0.17293 &  0.42495 &  1.02063 &  2.30254 \\
  \hline
o & LO  & 1.0786  & 2.7676 & 6.95356 & 16.21993 \\
  & NLO  &  0.35211 & 0.91745  & 2.29135 & 5.30139 \\
  & N3LO &  0.44408 & 1.16754  & 2.93643 & 6.82651 \\
\hline
\multicolumn{6}{c}{(B)}\\
Model&  order&  1.0\, {\rm fm}  & 1.5 \, {\rm fm} & 2.0\, {\rm fm} &  2.5\, {\rm fm}\\
\hline\hline
a & LO   & 1.793374 & 4.531530 & 11.44228 & 26.8957 \\
  & NLO  & 0.102262 & 0.267979 & 0.668900 & 1.54557 \\
  & N3LO & 0.098547 & 0.258262 & 0.644450 & 1.48855  \\
  \hline
b & LO   & -0.015077 & -0.036880 & -0.087577 & -0.19526 \\
  & NLO  &  0.28620  &  0.74440  &  1.85546  &  4.28638 \\
  & N3LO &  0.42761  &  1.11761  &  2.80328  & 6.50837   \\
  \hline
c & LO   & -2.130138 & -5.480962 & -12.69759 & -27.4026 \\
  & NLO  & -0.52108  & -1.35981  & -3.33139  &  -7.56723\\
  & N3LO & -0.030894 & -0.079822 & -0.196845 &  -0.45074\\
  \hline
d & LO   & -3.921656 & -11.04952 & -25.61489 & -54.3297 \\
  & NLO  & -0.63311  &  -1.58874 &  -3.78706 & -8.44497 \\
  & N3LO &  0.12387  &   0.30509 &   0.73318 & 1.65432 \\
  \hline
o & LO  & 1.0362  & 2.6637 & 6.69515 & 15.6184 \\
  & NLO  &  0.30552 & 0.79787  & 1.99382 & 4.6136 \\
  & N3LO &  0.39833 & 1.04955  & 2.64115 & 6.14099 \\
\end{tabular}
\end{ruledtabular}
\label{table:b416}
\end{center}
\end{table*}
In this section we report the results for the binding energies
of $^3$H, $^3$He, $^4$He, $^6$He, $^6$Li, $^{16}$O, $^{40}$Ca,
$^{48}$Ca, and $^{90}$Zr.  The calculations are carried out with $2N$
interactions up to N3LO in the $A\,$=$\,3$--6 systems, up to NLO in $^{16}$O,
and at LO only for the heavier nuclei with $A\ge 40$, and make use of
hyperspherical-harmonics (HH) methods in $A\le 6$ and auxiliary-field
diffusion Monte Carlo (AFDMC) methods for $A\ge 16$, see below.
Of course, a consistent study of nuclei must retain the complete
interaction at the different orders.  In the present work, which
deals primarily with the construction of $2N$
interactions, we
include the three-nucleon ($3N$) contact interaction at LO only, and
postpone the study of higher order $3N$ terms~\cite{Girlanda:2011fh}
to a subsequent work (a preliminary study of these higher order terms
can be found in Ref.~\cite{Girlanda:2018xrw}). At LO we consider
\begin{equation}
V_{\rm LO}=c_E\, \frac{f_\pi^4}{\Lambda_\chi}\,  \frac{(\hbar c)^6}{\pi^{3}\,R_{3}^6}\sum_{{\rm cyclic}\,ijk}{\rm e}^{-(r^2_{ij}+r^2_{jk})/R_{3}^2}\ ,
\end{equation}
where $\Lambda_\chi\,$=$\,1\,$GeV is the breaking scale of the theory and
$f_\pi\,$=$\,92.4$ MeV is the pion decay constant.  The LEC $c_E$
can be determined from a single three-nucleon data point for different choices of the range
$R_3$. Examples of this procedure can be found in Refs.~\cite{Kievsky:2015dtk,Kievsky:2018xsl,Gattobigio:2019omi} where 
correlations between the three-, four-, six-nucleon systems, and nuclear matter
have been analyzed.  Here, for each $2N$ model, we fix $c_E$ to reproduce $B({}^3{\rm H})=8.475\,$ MeV, 
for different choices of the cutoff $R_3$. The values so obtained are listed in
Table~\ref{table:b416}.

\subsection{Binding energies of $A\,$=3, 4, and 6 nuclei with HH methods}

The $^3$H, $^3$He, $^4$He, $^6$Li and $^6$He binding energies obtained with the different
$2N$ contact interactions are reported in Table~\ref{table:b34}.   As already noted, the
calculations have been carried out with the HH method, as described in
the recent reviews~\cite{Kievsky:2008es,Marcucci:2019hml} (and references therein).
These binding energies are expected to be accurate at the level of 1 keV and 10 keV
for the three- and four-nucleon systems, respectively.  
For the six-nucleon system the HH basis is largely degenerate requiring detailed studies. 
Accordingly,
the HH states are partitioned in different ``classes of convergence'' and within each of these
an extrapolation is made to estimate the missing energy. The estimates for all
classes of convergence are then added up to obtain the total extrapolated value for the binding energy.
A complete discussion of these aspects---in particular, the definition of classes of convergence,
and the extrapolation in each of these classes---can be found in Ref.~\cite{Gnech:2020qtt} for $^6$Li and in Ref.~\cite{Gnech:2021}
for $^6$He. The errors on the extrapolated energies are in general larger for $^6$He because
of the loosely bound structure and the slower convergence as compared to $^6$Li.
\begin{table*}[bth]
\caption{Binding energies (in MeV) corresponding to the $2N$ contact interaction models a-d and o,
obtained at LO, NLO, and N3LO with
the HH method for nuclei with mass number $A\,$=$\,3$, 4, and 6; the numbers in parentheses
for $A\,$=$\,6$ are estimates of extrapolation errors (see text).  The experimental
values are 8.48, 7.72, 28.3, 32.0, and 29.3 MeV for, respectively, $^3$H, $^3$He, $^4$He,
$^6$Li, and $^6$He.}
\begin{center}
\begin{ruledtabular}
\begin{tabular}{ccdddddd}
Model&  order& ^3{\rm H} &  ^3{\rm He} & ^4{\rm He} & ^6{\rm Li} & ^6{\rm He}\\
\hline
a & LO    &  10.705 & 9.917  &  40.89  & 46.71(3) & 43.03(7) \\
  & NLO   &  8.588 & 7.889  &  31.18   & 36.28(17)& 31.99(19)\\
  & N3LO  &  8.584 & 7.886  &  31.15   & 36.27(15)& 31.93(17)\\
\hline
b & LO    &  8.463 & 7.795  &  30.55  & 35.89(2) & 32.10(5) \\
  & NLO   &  8.790 & 8.084  &  32.49  & 37.29(7) & 33.15(9) \\
  & N3LO  &  8.964 & 8.249  & 33.36   & 38.73(6) & 34.51(8) \\
\hline
c & LO    &  7.066 & 6.483  & 24.29 & 29.20(2) & 25.52(4)  \\
  & NLO   &  7.967 & 7.307  & 28.13 & 32.00(10)& 27.66(14) \\
  & N3LO  &  8.443 & 7.757  & 30.08 & 39.10(15)& 33.97(10) \\
\hline
d & LO    &  6.136 & 5.617  &  20.21  & 24.78(2) & 21.24(3)\\
  & NLO   &  7.941 & 7.299  &  28.29  & 32.54(5) & 28.35(5)\\
  & N3LO  &  8.589 & 7.912  &  31.02 &  50.10(3) & 44.26(3)\\
\hline
o & LO      &  9.696  & 8.958 &  36.88 & 42.27(4) & 37.71(8) \\
  & NLO     &  8.816  & 8.107 &  32.41 & 37.38(12)& 33.14(16)\\
  & N3LO    &  8.937  & 8.221 &  33.17 & 38.68(11)& 34.30(11)\\
\end{tabular}
\end{ruledtabular}
\label{table:b34}
\end{center}
\end{table*}


We find that at LO there is a large spread in the results, reflecting a large dependence on
the cutoffs.  The three-nucleon binding energies vary by more than 4 MeV, whereas the
spread in the four- and six-nucleon binding energies exceeds 20 MeV.  This large variation
as a function of the cutoffs is related to the Thomas collapse phenomenon~\cite{Thomas:1935zz}:
as the range of the interaction is reduced these systems tend to become more and more bound.
This is especially apparent in the limiting case in which the interaction is of zero-range,
corresponding to the limiting case of the LO interaction. As is apparent from Table~\ref{table:b34},
the dependence on the cutoffs is much less drastic for the NLO and N3LO interactions.
However, the $^6$Li and $^6$He results show a peculiar behavior, in that at N3LO the spread
is relatively small, about 3 MeV, for the a, b, c, and o models; on the other hand, model d seems to be an
outlier and yields large binding energies, when compared to those of the other models.
Lastly, it should be noted that $^6$He is found to be bound with all models, except with model c at NLO.

\begin{figure}[bth]
\includegraphics[width=6.65in]{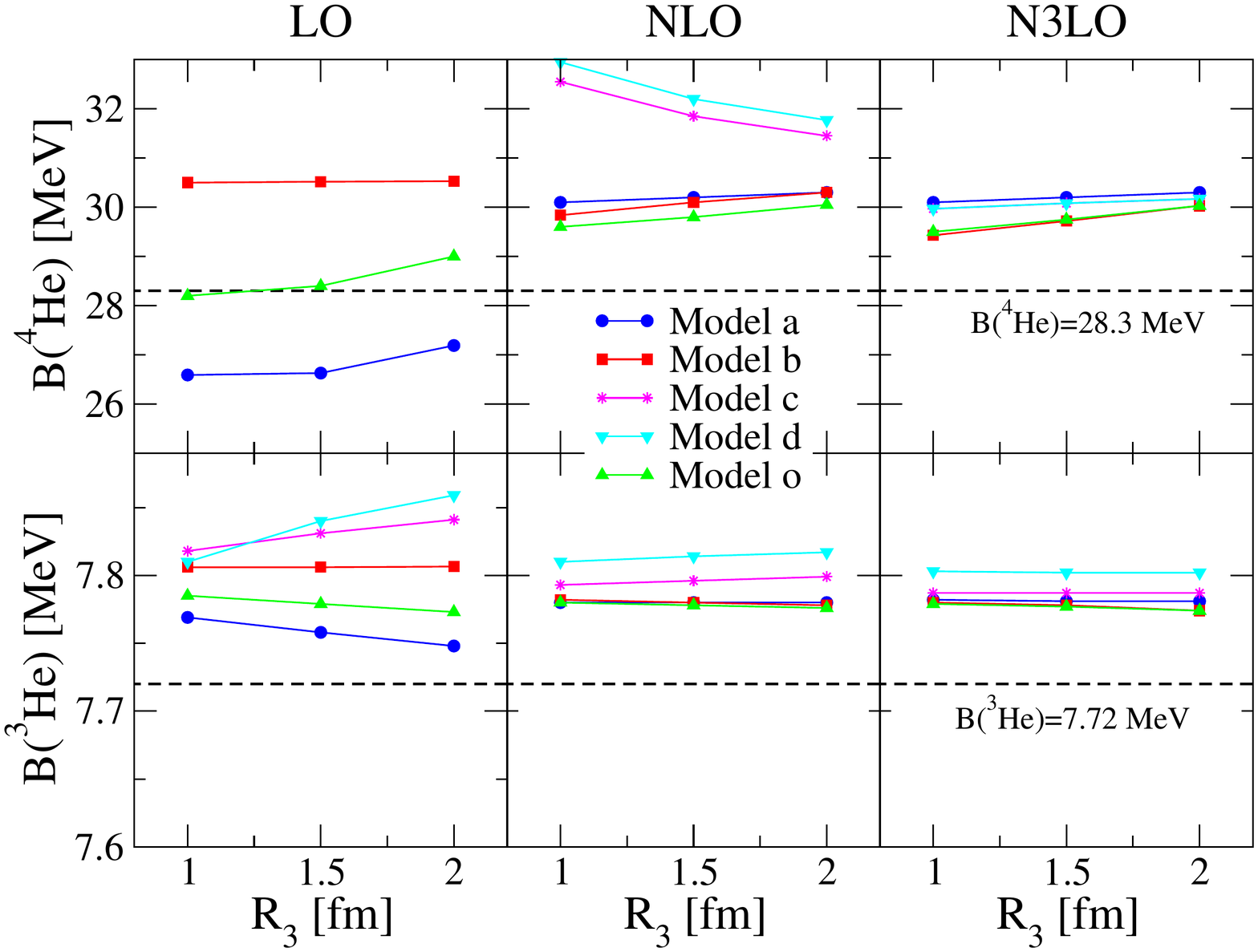}
\caption{(Color online). Binding energies of $^3$He (lower sub-panels) and $^4$He (upper sub-panels) 
with the inclusion of the LO $3N$ contact interaction corresponding to set B, determined by fitting the $^3$H binding energy,
as a function of the cutoff $R_3$.  The left, center, and right panels refer to the LO, NLO and N3LO $2N$
contact interactions (each in combination with the LO $3N$ contact interaction), while circles, squares, stars, triangles down, triangles up  correspond to
models, a-d and o, respectively.  The lines are only drawn to guide the eye. For $^4$He the LO binding energies obtained with models c and d are out of scale
(too large) and are not shown.  The dash lines indicate the experimental values;
note the different energy scales adopted for $^3$He and $^4$He.}
\label{fig:r3}
\end{figure}

Next, we include the $3N$ contact interaction discussed earlier in the binding energy calculations
of the $A\,$=$\, 3$ and 4 nuclei.  The results are summarized in Fig.~\ref{fig:r3}.
The left, center, and right panels present the binding energies obtained with the
$2N$ contact interactions at LO, NLO, and N3LO (each including the full
electromagnetic interaction and hence each in combination with
the LO $3N$ contact interaction corresponding to set B of $c_E$ values), whereas the different symbols in
each panel correspond to the five possible choices of $2N$ cutoffs. The lower and
upper sub-panels show the $B({}^3{\rm He})$ and $B({}^4{\rm He})$ results (note the energy scales).
The two dashed lines indicate the experimental values, $B({}^3{\rm He})\,$=$\,7.72\,$MeV
and $B({}^4{\rm He})\,$=$\,28.3\,$MeV.  The constraint
$B({}^3{\rm H})\,$=$\,8.475\,$MeV is verified by all models by construction. The figure shows
that at LO a fine tuning of the cutoffs in the $2N$ interaction could be used to reproduce
$B({}^4{\rm He})$, since models a and b are just below and above the experimental value; such fine tuning is in fact
achieved with model o.
Increasing the order of the expansion, at NLO and N3LO, leads to much more stable results,
clustering around $7.78\,$ MeV for $B({}^3{\rm He})$ and around $30\,$ MeV 
for $B({}^4{\rm He})$.  We expect these binding energies to be correctly
reproduced by including higher-order terms in the $3N$ interaction.

\subsection{$^{16}$O nucleus with AFDMC methods}
The auxiliary-field diffusion Monte Carlo (AFDMC) method~\cite{Schmidt99} is used to study nuclei with $A>6$ nucleons---see Ref.~\cite{Gandolfi20} for a recent review. The AFDMC method uses imaginary-time projection techniques to filter out the ground-state of the system starting from a suitable trial wave function,
$|\Psi_0\rangle = e^{-(H-E_0)\tau} |\Psi_T\rangle$,
and exhibits a favorable polynomial scaling with the number of nucleons, which is made possible by the use of a single-particle spin-isospin basis. This representation is preserved during the imaginary-time evolution by using Hubbard-Stratonovich transformations to linearize the quadratic spin-isospin operators entering the short-time propagator. Applying these transformation to treat the isospin-dependent spin-orbit term, implicit in the
$v_{\rm NLO}^{\rm CI}$ interaction, involves non-trivial difficulties. To circumvent them, we perform the imaginary-time propagation with a modified interaction, as described below.
The $v_{\rm NLO}^{\rm CI}$ interactions reads
\begin{equation}
v_{\rm NLO}^{\rm CI}=\sum_{l=1}^6 v^l(r)\, O_{12}^l+[v^b(r)+v^{b\tau}(r)\, {\bm \tau}_1\cdot{\bm \tau}_2]\,
{\bf L}\cdot{\bf S} \ ,
\end{equation}
where, referring to Appendix~A, the radial functions of the spin-orbit components
are defined as
\begin{eqnarray}
v^b(r)&=&-\frac{C_7}{r}\, \frac{3\,C^{(1)}_1(r)+C_0^{(1)}(r)}{4} \ , \\
v^{b\tau}(r)&=&-\frac{C_7}{r}\, \frac{ C^{(1)}_1(r)-C_0^{(1)}(r)}{4}\ ,
\end{eqnarray}
with $C^{(1)}_T(r)$ denoting the derivative of the Gaussian cutoff in isospin channel $T$.
The isospin dependence of the spin-orbit term comes on account of the
fact that $C_0^{(1)}(r)$ and $C_1^{(1)}(r)$ have different ranges.
The modified NLO interaction is defined as
\begin{equation}
v^{\rm CI\, \prime}_{\rm NLO}(\alpha) = \sum_{1=1}^6 v^l(r)\, O^l_{12} +
[ v^b(r)+\alpha\, v^{b\tau}(r)]\,{\bf L}\cdot {\bf S}\ .
\end{equation}
The imaginary time propagation is performed with this modified interaction. 
The expectation values of both $v^{\rm CI\, \prime}_{\rm NLO}(\alpha)$ and the original $v_{\rm NLO}^{\rm CI}$
are then evaluated, and the parameter $\alpha$ is adjusted so as to make these expetaction
values the same within statistical errors. Note that a similar procedure has been adopted in Refs.~\cite{Lonardoni18, Lonardoni18b} to include the commutator term of the three-body chiral interaction. 

The trial wave function is expressed as a product of a long-range Slater determinant of single-particle orbitals and a correlation factor, $|\Psi_T\rangle = F\,|\Phi\rangle$. Since the NLO Hamiltonian contains both tensor and spin-orbit terms, we consider a trial wave function that includes linear spin-isospin dependent correlations~\cite{Gandolfi13}
\begin{equation}
F = \prod_{i<j<k} \left[1 +\sum_{\rm cyc} u_{\rm 3b}(r_{ij})u_{\rm 3b}(r_{ik})\right]\left[1 + \sum_{i<j} \sum_{l=2}^6 u^l(r_{ij}) O^l_{ij}\right] \, \prod_{i<j} f^c(r_{ij})
\label{eq:trial_wf}
\end{equation}
where the spin-isospin operators are defined in Eq.~\eqref{eq:vr}. The functions $u^l(r)$ are characterized by a number of variational parameters~\cite{Gandolfi20}, which are determined by minimizing the two-body cluster contribution to the energy per particle of nuclear matter at saturation density.  On the other hand, the function $u_{\rm 3b}(r)$ associated with the correlations induced
by the (LO) $3N$ contact interaction, and the function $f^c(r)$
are parametrized in terms of cubic splines. The variational parameters are the values of $u_{\rm 3b}(r)$ and $f^c(r)$ at the grid points, plus the value of their first derivatives at $r\,$=$\, 0$. The optimal values of the variational parameters  are found employing the linear optimization method~\cite{Contessi17}, which typically converges in $\approx 20$ iterations. When solving the LO Hamiltonian, which does not contain tensor or spin-orbit terms, we drop the spin-isospin dependent correlations in Eq.~(\ref{eq:trial_wf}). This simplified ansatz is consistent with that adopted in Ref.~\cite{Contessi17}, and allows us to compute nuclei as large as $^{90}$Zr with multiple LO Hamiltonians. 

In Table~\ref{table:afdmc34} we report the AFDMC binding energies of $^3$H and $^4$He obtained in the constrained-path approximation using the linearized spin-isospin correlations of Eq.~\eqref{eq:trial_wf} and compare them with the HH results. Since in the AFDMC the electromagnetic interaction only includes the Coulomb repulsion between finite-size (rather than point-like) protons, for a more meaningful comparison, the HH binding energies are also obtained with this approximation; hence these energies are slightly different from those of Table~\ref{table:b34} which retain the full electromagnetic interaction. The AFDMC and HH results for $^3$H are in excellent agreement with each other: the largest discrepancy between the two methods is $\approx 0.05$ MeV for model c at NLO; differences between AFDMC and HH results are much smaller for all the other models. A similar trend is observed for the $^4$He nucleus; the AFDMC and HH energies are quite close; the largest discrepancy, $\approx 0.13$ MeV, is again observed for model a at NLO, and is smaller for all other models we analyzed. Some of these differences can be ascribed to a combination of the constrained-path approximation employed in the AFDMC, the approximate treatment of the isospin-dependent spin-orbit term of the interaction, and the convergence of the HH basis expansion. 

The binding energies of $\,^4$He and $\,^{16}$O at LO, NLO, and N3LO for selected $2N$ models and including the $3N$ interaction are listed in Table~\ref{table:b416ab}. The agreement between HH and AFDMC calculations
of $^4$He---the latter reported in square brackets---remains excellent even when the $3N$ interaction is included in the Hamiltonian. We note that neglecting the $3N$ interaction always yields too large a binding energy in $^{16}$O, even when the $^4$He is underbound. On the other hand, fixing the $3N$ interaction to
reproduce the $^3$H binding energy leads to a sizable cutoff dependence of our results, regardless of the $2N$ interaction model considered. In general, a larger cutoff $R_3$ corresponds to a lesser bound $^{16}$O, as the repulsive term becomes long-ranged and affects triplets of nucleons belonging to different $\alpha$-like clusters. In this regard, we observe that in some cases the AFDMC binding energies of $^{16}$O are smaller than four times that of $^{4}$He. Although a fully clusterized wave function can be obtained as done in Ref.~\cite{Contessi17}, in this work we use confining single-particle orbitals that effectively prevent the nucleons from diffusing far from the center of mass of the system. Finally, we refrain from carrying out AFDMC calculations of $^{16}$O for models c and d, since for these the LEC $c_E$ is negative and the corresponding $3N$ interaction would therefore lead to large additional binding for the already overbound results predicted by the $2N$ models alone.  

In Table~\ref{table:b416a} we list the binding energies of selected light- and medium-mass nuclei at LO computed using the HH and AFDMC methods. We observe a similar trend as in Table~\ref{table:b416ab}, with a sizable dependence of the results on the cutoff $R_3$. Overbinding or underbinding in $^{16}$O carries over in heavier nuclei. On a positive note, models a and o for $R_3=1.0$ fm provide a satisfactory description of $^{16}$O and are also able to reproduce fairly well the binding energies of heavier systems. It would be
interesting to fit $c_E$ by reproducing $^{16}$O, as done, for instance in Ref.~\cite{Ekstrom15}, and study the behaviour of such a Hamiltonian in lighter and heavier nuclei.

\begin{table*}[h]
\caption{Binding energies (in MeV) of $^3$H and $^4$He obtained with LO and NLO $2N$ interactions using the AFDMC method in the constrained-path approximation are compared to corresponding HH results. Estimated statistical errors in the AFDMC calculations are in parentheses. Note that the electromagnetic interaction
only includes the Coulomb repulsion between finite-size (rather than point-like)
protons.}
\begin{center}
\begin{ruledtabular}
\begin{tabular}{cc cccc}
Model&  order&  $B({}^3{\rm H})$ & $B({}^3{\rm H})$ [HH] & $B({}^4{\rm He})$ & $B({}^4{\rm He})$ [HH]\\
\hline
a & LO    & 10.75(2) & 10.756 & 41.10(5) & 41.10 \\
  & NLO   & 8.64(1)  & 8.639  & 31.50(2) & 31.37 \\
\hline
b & LO    & 8.47(1) &  8.498  & 30.71(1) & 30.69 \\
  & NLO   & 8.82(1) &  8.839  & 32.75(1) & 32.68 \\
\hline
c & LO    & 7.07(1) &  7.093  & 24.40(1) & 24.40\\
  & NLO   & 7.96(2) &  8.013  & 28.33(2) & 28.31 \\
\hline
d & LO    & 6.14(1) &  6.158  & 20.30(2) & 20.30\\
  & NLO   & 7.96(2) &  7.981  & 28.45(1) & 28.44\\
\hline
o & LO    & 9.71(1) &  9.744  & 37.08(3) & 37.07 \\
  & NLO   & 8.85(1) &  8.867  & 32.66(3) & 32.60 \\
\end{tabular}
\end{ruledtabular}
\label{table:afdmc34}
\end{center}
\end{table*}
\begin{table*}[h]
\caption{Binding energies (in MeV) of $\,^4$He and $\,^{16}$O at LO, NLO, and N3LO obtained with
selected combinations of contact $2N$+$3N$ interaction models, and corresponding to different
cutoffs in the $3N$ interaction, as reported in Table~\ref{table:b416}. Note that in these calculations
we have retained in $v^{\rm EM}$ only the Coulomb interaction between protons
(albeit accounting for their finite size). 
Consequently, we have used the $c_E$ values reported in part (A) of Table~\ref{table:b416}. The $A\,$=$\,4$ results are
calculated with both the HH method and, in square brackets, the AFDMC method in the constrained-path
approximation; the latter method is used in the $A\,$=$\,16$ calculations. }
\begin{center}
\begin{ruledtabular}
\begin{tabular}{cc cc cc cc cc}
Model&  order&  $A=4$ &  $A=16$ & $A=4$ &  $A=16$ & $A=4$ &  $A=16$ &  $A=4$ &  $A=16$  \\
\hline
&&\multicolumn{2}{c}{no $3N$} &
\multicolumn{2}{c}{$R_3=1.0$ fm} &
\multicolumn{2}{c}{$R_3=1.5$ fm} &
\multicolumn{2}{c}{$R_3=2.0$ fm}\\
a & LO   & 41.10 & 355.7(2) & 26.59 [26.57(2)] & 111.6(3)  & 26.63 [26.62(2)] & 76.5(8) & 27.19 [27.19(1)] & 69.1(9)\\
  & NLO  & 31.37 & 424.7(4) & 30.08 [30.19(2)] & 260.7(8)  & 30.20 [30.31(2)]  & 243.2(4) & 30.30 [30.39(3)] & 243.6(4)\\
  & N3LO & 31.15 & & 30.08 &   & 30.20 & & 30.30 &\\
  \hline
b & LO   & 30.69 & 262.8(9) & 30.50 [30.51(2)] & 260.5(9) & 30.52 [30.52(2)] & 251.0(6)& 30.53 [30.53(2)] & 249.3(8)\\
  & NLO  & 32.68 & 367.2(3) & 29.84 [29.89(2)] & 194.8(6) & 30.08 [30.12(3)] & 163.8(5) & 30.29 [30.36(2)]  &133.7(9)\\
  & N3LO & 33.36 & & 29.43 &   & 29.72 & & 30.03 &\\
  \hline
c & LO   & 24.39 & 206.9(8) & 47.47 &  & 36.98 & & 34.67 & \\
  & NLO  & 28.31 & 317.7(3) & 32.55 &  & 31.85 & & 31.45 & \\
  & N3LO & 30.26 & & 30.13 &  & 30.14 & & 30.15 & \\
  \hline
d & LO   & 20.29 & 170.1(4) & 139.1 &  & 49.46 &  & 40.24 & \\
  & NLO  & 28.44 & 229.0(9) & 32.95 &  & 32.20 &  & 31.77  &\\
  & N3LO & 20.29 & & 29.97 &  & 30.08 &  & 30.17 &\\
  \hline
o & LO   & 37.07 & 278.9(9) & 28.18 [28.23(2)] & 133.4(4) & 28.49 [28.49(1)] & 96.6(4) & 29.00 [29.02(2)]& 69.7(5)\\
  & NLO  & 32.60 & 436.8(9) & 29.62 [29.69(2)] & 200.1(3)& 29.85 [29.89(2)] & 157.0(3) & 30.06 [30.14(2)] & 125.8(5)\\
  & N3LO & 33.17 & & 29.47  &   & 29.75 & & 30.03 &\\
\end{tabular}
\end{ruledtabular}
\label{table:b416ab}
\end{center}
\end{table*}

\begin{table*}[h]
\caption{Binding energies (in MeV) of light- and medium-mass nuclei at LO
predicted by
selected combinations of contact $2N$+$3N$ interaction models, corresponding to different
cutoffs in the $3N$ interaction as reported in Table~\ref{table:b416}. 
The calculations retain in $v^{\rm EM}$ only the Coulomb repulsion
between finite-size (rather than point-like) protons.
Consequently, the $c_E$ values reported in part (A) of Table~\ref{table:b416}
have been used for the (LO) $3N$ interaction.
The $A$=4--6
and $A\ge 16$ results are obtained, respectively, with the HH method and
AFDMC method in the constrained-path
approximation. }
\begin{center}
\begin{ruledtabular}
\begin{tabular}{c|ccc|ccc|ccc|c}
Nucleus&    a & b & o &  a  & b  &  o & a & b& o &Exp.\\
\hline
&\multicolumn{3}{c}{$R_3=1.0$ fm} &
\multicolumn{3}{c}{$R_3=1.5$ fm} &
\multicolumn{3}{c}{$R_3=2.0$ fm} & \\
\hline
$^4$He &  26.59 & 30.50  & 28.18   & 26.63 & 30.52 & 28.49 &27.19  & 30.53 & 29.00 & 28.30\\
$^6$Li &  28.55(2)& 35.62(1) & 30.77(1)  & 26.70(3) & 35.59(1) & 29.51(1) & 25.07(5) & 35.57(1) & 28.28(1) & 31.99\\
$^6$He &  25.73(2)& 31.89(3) & 27.26(3)  & 23.96(5) & 31.87(3) & 26.20(6) & 22.46(9) & 31.85(3) & 25.22(9) & 29.27\\
$^{16}$O &111.6(3)  &260.5(9)  &133.4(4)   & 76.5(8) & 251.0(6) & 96.6(4) &69.1(9)  & 249.3(8) & 69.7(5)& 127.62\\
$^{40}$Ca & 297.6(5) & 1463.0(9) & 395.2(8)  & 147.2(8) & 1491.7(9) & 207.2(9) & 234(2)  & 1462(1) & 120.7(8) & 343.05\\
$^{48}$Ca & 332.4(5) & 1873.3(5)   & 446.5(9)  & 159.3 (6) & 1927(2)   &  225.9(8) & 161(9) & 1874(1) & 130(2) & 416.00\\
$^{90}$Zr  & 654(2) & 6511(9)    &937(6)  & 216.9(8) & 6866(9) &  392(2) & - & 6363(8) & - & 783.90\\
\end{tabular}
\end{ruledtabular}
\label{table:b416a}
\end{center}
\end{table*}

\section{Summary and conclusions}
\label{sec:concl}

The present work represents the first phase in a program we envision,
aimed at establishing whether the energy spectra of, and electroweak 
transitions between, low-lying states of nuclei can be understood on the
basis of nuclear interactions and electroweak currents, derived in an EFT
formulation where pion degrees of freedom have been integrated out.
Specifically, this first phase has dealt with: (i) the construction of $2N$
contact interactions at LO, NLO, and N3LO that are local in configuration
space and therefore suitable for implementation in  quantum Monte Carlo
calculations; (ii) the determination of a $3N$ contact interaction at LO with
the single LEC entering at this order fixed to reproduce the tritium binding
energy in essentially exact HH calculations; (iii) the extension of the
AFDMC method, so as to be capable to handle approximately but reliably
tensor and spin-orbit components (with and without isospin
dependence) in the $2N$ interactions; (iv) a fairly complete study
(albeit not a fully consistent one from a power counting perspective)
of the ground-state binding energies of light nuclei with mass number
up to $A\,$=$\,16$ with Hamiltonians based on $2N$ interactions of
increasing order but a $3N$ interaction included only at LO; (v) an
initial set of AFDMC calculations of the binding energies of nuclei
with $A\ge 40$ based on Hamiltonians including the contact $2N$ and
$3N$ interactions at LO only.

The fits to the $2N$ scattering database (including the deuteron binding
energy) have been restricted up to lab energies of 15 MeV at NLO
and 25 MeV up to N3LO.  Despite the (apparent) flexibility afforded
by the 25 LECs (20 in the charge-independent sector
and 5 more in the charge-dependent one) present in the interaction at N3LO,
it has proven rather difficult to extend the fits much beyond 25 MeV,
while at the same time maintaining a $\chi^2/{\rm datum} \lesssim 2$.
This may not be surprising, given that in the present EFT formulation
the expansion parameter $Q/\Lambda$ with $Q$ and $\Lambda$ being
taken, respectively, as the relative momentum and pion mass, is $\approx 0.78$
at a lab energy of 25 MeV.

A different but potentially related issue is the presence of local minima in
the $\chi^2$-minimization.
There might be more efficient and effective means, like those based on Bayesian methods or
machine-learning techniques, to explore the parameter space than standard
optimization packages, such as POUNDERS employed in the present work.
An exploratory investigation along these lines is in progress.   

In a point of departure from the standard approach, we found it helpful
to have LO interactions with projections only in the spin singlet $T\,$=$\,1$
and spin triplet $T\,$=$\,0$ channels, by fixing the associated
$C_{01}$ and $C_{10}$ LECs so as to reproduce, respectively, the (large) singlet
scattering length and deuteron binding energy.  These interactions vanish
in odd partial waves (in particular, P-waves) that are unconstrained
by data (at this order), and, as a consequence, significantly improve the LO
description of ground-state energies in $A\ge 6$ nuclei. In this respect,
we note that a fine tuning of the LECs $C_{01}$ and $C_{10}$ and corresponding
cutoffs $R_1$ and $R_0$ leads to a LO interaction (model o), that correctly
reproduces the $np$ effective range expansions in spin-singlet and spin-triplet
channels. 

All the N3LO $2N$ interactions overestimate the $^3$H binding energy except
for model c, which leads to an under-binding of about of 40 keV (see
Table~\ref{table:b34}).  As a result, the LEC $c_E$ in the $3N$
contact interaction accompanying each of these (N3LO) $2N$ interactions is positive
and therefore produces a repulsive contribution for all models except model c,
see Table~\ref{table:b416} part B.  However, when $\alpha^2$
corrections are ignored in the electromagnetic interaction, the N3LO model c 
also leads to over-binding in $^3$H and hence to a repulsive $3N$ interaction,
see again Table~\ref{table:b416} but now part A.  The results in Table~\ref{table:b34}
also indicate that the LO and NLO $2N$ interactions typically overbind (underbind)
$^3$H when the harder (softer) cutoffs, that is, smaller (larger) values for
$R_0$ and $R_1$ are adopted.  For the $2N$ interactions with
the softer cutoffs the need to have an attractive $3N$ contribution ($c_E <0$)
in order to reproduce the experimental value of the $^3$H binding energy
proves catastrophic in larger nuclei, for example, by wildly over-predicting
the $^{16}$O binding energy.   Indeed, these $2N$ and $3N$ models have
not been considered in calculations of nuclei with $A\ge 16$.

\begin{figure}[bth]
\includegraphics[width=7.65in]{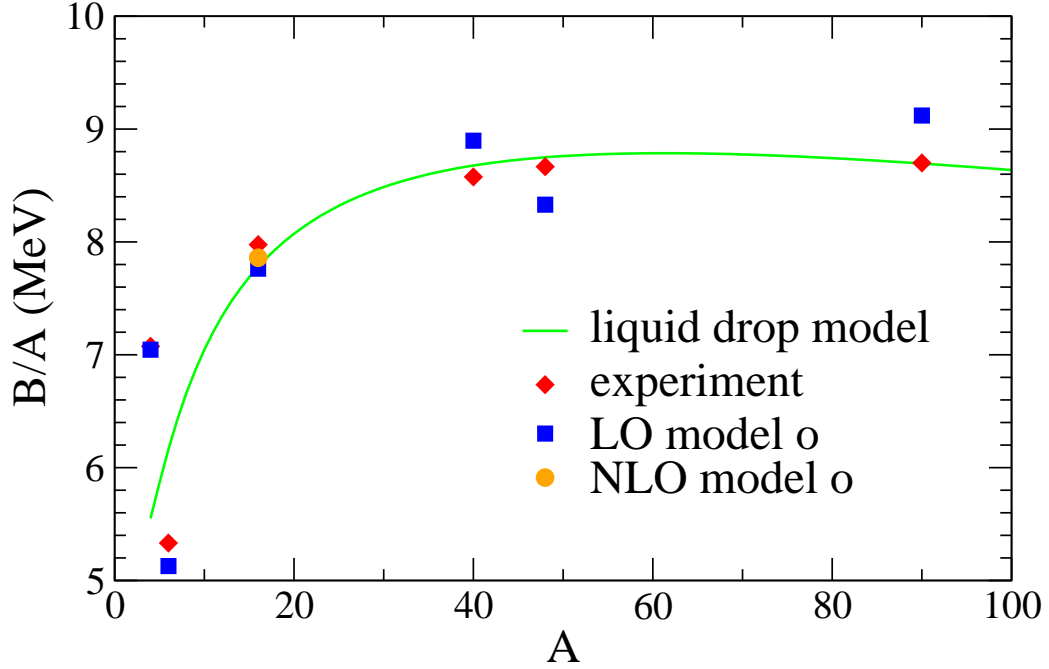}
\caption{Binding energy per particle as a function of the
atomic number $A$. The liquid drop model results, along with specific experimental
values (red symbols) in the cases of $A=4,16,40,48,90$, are shown.  Predictions
obtained in the present work with the $2N$ interaction model o at either
LO or NLO in combination with the $3N$ interaction at LO, are indicated by the blue (orange) symbols.}
\label{fig:ldm}
\end{figure}
There is a large dependence of the calculated binding energies,
particularly in $A\ge 16$, on the cutoff of the $3N$ interaction,
the softer cutoff $R_3\,$=$\, 2.0$ fm generally yielding binding
energy values closer to experiment, see Table~\ref{table:b416a}.
In one case, model o at LO with $R_3\,$=$\,1.0$ fm and
at NLO with $R_3\,$=$\,2.0$ fm, theory
is (miraculously, perhaps) within less than 2\% of experiment.
As matter of fact, the overall picture of nuclear ground-state
spectra that emerges from the LO and NLO Hamiltonians
corresponding to model o (that is, the model o $2N$ interaction
at either LO or NLO in combination with the LO $3N$ interaction
having cutoff $R_3\,$=$\,1$ fm and $2$ fm, respectively) is summarized in Fig.~\ref{fig:ldm},
where the predicted binding energies per nucleon are shown
as function of the mass number $A$ (only LO results are available
in $A\ge 40$), and are compared to experimental and liquid-drop
mass formula values.

The optimization of the two-body ($R_0$ and $R_1$) and three-body ($R_3$)
cutoffs has different motivations. As we have already mentioned, the optimization
of $R_0$ and $R_1$ leads to the correct description of effective range parameters,
and to an appreciable improvement in the $\chi^2$ values obtained in fits to the $2N$ database. 
On the other hand, the optimization of $R_3$ is aimed at providing a satisfactory description of
nuclear binding energies. We have observed that within model o, when $R_3$ is set to reproduce
reasonably well these energies in the mass range $A\le 16$, the energies for $40\le A\le 90$ are
also reasonably well reproduced.  We conclude that this parameter can be used to balance approximately
the repulsion-attraction ratio of the nuclear Hamiltonian. A similar situation has been recently observed
in the case of bosonic helium clusters~\cite{kievsky:2017,kievsky:2020}.

Because of the more complicated operator structure of the
$2N$ interactions at N3LO (in particular the presence
of ${\bf L}^2$ and $({\bf L}\cdot{\bf S})^2$ terms), these have
yet to be implemented in the AFDMC method, and therefore
in the present study no N3LO results are reported for $A\ge 16$.
However, even within the context of calculations based on the NLO
$2N$ interactions, there are sub-leading $3N$ contact interactions,
suppressed by $Q^2/\Lambda^2$ relative to the LO ones, that need
to be accounted for.  These terms have a rich operator structure
including central, tensor and spin-orbit-like components, but involve a relatively
large number (13) of unknown LECs~\cite{Girlanda:2011fh}.  Arguments
based on the large $N_c$ expansion allow one
to reduce the number of LECs and associated operator structures by 
ranking their relative importance~\cite{Girlanda:2018xrw}.  Nevertheless, the problem remains
of how best to determine these LECs.  Two alternative strategies are to
constrain them by fitting $3N$ scattering data (cross
sections and polarizations) at low energies or by reproducing
the energies of low-lying states of selected light nuclei. Both alternatives
should be investigated.

\acknowledgments
This research is supported by the U.S. Department of Energy, Office
of Nuclear Science, under contracts DE-AC05-06OR23177 (A.G. and R.S.), DE-AC02-06CH11357 (A.L.), 
and the U.S. Department of Energy through the FRIB Theory Alliance award DE-SC0013617 (M.P.). A.L. is also supported by U.S. Department of Energy Early Career Research Program award. The calculations were made possible by grants of computing time
from the Argonne Laboratory Computing Resource Center (LCRC), the Argonne
Leadership Computing Facility (ALCF) via the
2020/2021 ALCC grant ``Chiral Nuclear Interactions from Nuclei to Nucleonic Matter"
for the project ChiralNuc, and the National Energy Research Supercomputer Center (NERSC).
Computing time on the MARCONI
supercomputer at CINECA was also utilized for some of the $A\,$=$\,6$ and $A\,$=$\,16$ calculations.
\appendix

\section{Configuration-space representation of the interactions}
\label{app:a1}
The coordinate-space representation of a generic (regularized) term
$O({\bf K},{\bf{k}})$ follows from
\begin{eqnarray}
O({\bf r})=\int \frac{d{\bf k}}{(2\pi)^3}
\int \frac{d{\bf K}}{(2\pi)^3} \, e^{i\,{\bf k}\cdot ({\bf r}^{\prime}+{\bf r})/2}
\,O({\bf K},{\bf{k}})\,e^{i\,{\bf K}\cdot ({\bf r}^{\prime}-{\bf r})}\ ,
\label{eq:ft}
\end{eqnarray}
where ${\bf r}$ is the relative position and ${\bf K}
\longrightarrow {\bf p} =-i\, {\bm \nabla}^\prime\delta({\bf r}^\prime-{\bf r})$,
the relative momentum operator.  For the momentum-space operator structures
present in the contact interactions we find:
\begin{eqnarray}
1 &\longrightarrow& C(r) \ , \\
k^2&\longrightarrow& - C^{(2)}(r)-\frac{2}{r}\, C^{(1)}(r)
\ ,\\
k^4 &\longrightarrow& C^{(4)}(r)+\frac{4}{r}\,C^{(3)}(r) \ ,\\
S_{12}({\bf k})&\longrightarrow&-\left[ C^{(2)}(r)-\frac{1}{r}\, C^{(1)}(r)\right] S_{12} \ , \\
i\, {\bf S} \cdot \left( {\bf K}\times{\bf k}\right) &\longrightarrow& -\frac{1}{r} \,
C^{(1)}(r)\,{\bf L}\cdot{\bf S} \ , \\
 \label{eq:eb8}
 \left( {\bf K}\times{\bf k}\right)^2 &\longrightarrow&-\frac{1}{r^2}
 \left[ C^{(2)}(r)-\frac{1}{r}\, C^{(1)}(r)\right] {\bf L}^2
+\cdots  \ , \\
 \label{eq:eb9}
  \left[{\bf S}\cdot\left( {\bf K}\times{\bf k}\right)\right]^2&\longrightarrow&
  -\frac{1}{r^2}
 \left[ C^{(2)}(r)-\frac{1}{r}\, C^{(1)}(r)\right] \left({\bf L}\cdot {\bf S}\right)^2
 +\cdots \ ,
\end{eqnarray}
where $C(r)$ is defined in Eq.~(\ref{eq:e2.6}) and
\begin{equation}
C^{(n)}(r) =\frac{d^n C(r)}{dr^n} \ .
\end{equation}
Note that in Eqs.~(\ref{eq:eb8}) and~(\ref{eq:eb9}) only the terms proportional
to ${\bf L}^2$ and $({\bf L}\cdot{\bf S})^2$ are retained; the $\cdots$ represent
additional terms which either involve terms quadratic in the relative momentum
operator or give rise to structures already accounted for.
Using the above expressions, the functions $v^{l}(r)$ for the CI terms are obtained as
\begin{eqnarray}
v^c(r)&=& v^c_{\rm LO}(r)+C_1\left[- C^{(2)}(r)-\frac{2}{r}\, C^{(1)}(r)\right]+
D_1\left[C^{(4)}(r)+\frac{4}{r}\,C^{(3)}(r)\right]\ , \\
v^\tau(r)&=&v^\tau_{\rm LO}(r)+ C_2\left[- C^{(2)}(r)-\frac{2}{r}\, C^{(1)}(r)\right]+
D_2\left[C^{(4)}(r)+\frac{4}{r}\,C^{(3)}(r)\right]\ , \\
v^\sigma(r)&=&v^\sigma_{\rm LO}(r)+C_3\left[- C^{(2)}(r)-\frac{2}{r}\, C^{(1)}(r)\right]+
D_3\left[C^{(4)}(r)+\frac{4}{r}\,C^{(3)}(r)\right]\ , \\
v^{\sigma\tau}(r)&=& v^{\sigma\tau}_{\rm LO}(r)+C_4\left[- C^{(2)}(r)-\frac{2}{r}\, C^{(1)}(r)\right]+
D_4\left[C^{(4)}(r)+\frac{4}{r}\,C^{(3)}(r)\right]\ , \\
v^{t}(r)&=& -C_5\left[C^{(2)}(r)-\frac{1}{r}\, C^{(1)}(r)\right]+D_5\left[
C^{(4)}(r)+\frac{1}{r}C^{(3)}(r)-\frac{6}{r^2}C^{(2)}(r)+\frac{6}{r^3}
C^{(1)}(r)\right]\ , \\
v^{t\tau}(r)&=& -C_6\left[C^{(2)}(r)-\frac{1}{r}\, C^{(1)}(r)\right]+D_6\left[
C^{(4)}(r)+\frac{1}{r}C^{(3)}(r)-\frac{6}{r^2}C^{(2)}(r)+\frac{6}{r^3}
C^{(1)}(r)\right]\ , \\
v^{b}(r)&=& -C_7\frac{1}{r}C^{(1)}(r)+D_7\left[\frac{1}{r}C^{(3)}(r)+2\,\frac{1}{r^2}C^{(2)}(r)-\frac{2}{r^3}C^{(1)}(r)\right]\ , \\
v^{b\tau}(r)&=& D_8\left[\frac{1}{r}C^{(3)}(r)+2\,\frac{1}{r^2}C^{(2)}(r)-\frac{2}{r^3}C^{(1)}(r)\right]\ , \\
v^{bb}(r)&=&-D_9\frac{1}{r^2}
 \left[ C^{(2)}(r)-\frac{1}{r}\, C^{(1)}(r)\right] \ , \\
v^{q}(r)&=&-D_{10}\frac{1}{r^2}
 \left[ C^{(2)}(r)-\frac{1}{r}\, C^{(1)}(r)\right] \ , \\
v^{q\sigma}(r)&=& -D_{11}\frac{1}{r^2}
 \left[ C^{(2)}(r)-\frac{1}{r}\, C^{(1)}(r)\right]\ , 
 \end{eqnarray}
 and those for the CD ones are obtained as
 \begin{eqnarray}
v^{T}(r)&=& C_0^{\rm IT}\,C(r)+C_1^{\rm IT}\left[- C^{(2)}(r)-\frac{2}{r}\, C^{(1)}(r)\right]\,\ , \\
v^{\sigma T}_{S}(r)&=& C_2^{\rm IT}\left[- C^{(2)}(r)-\frac{2}{r}\, C^{(1)}(r)\right]\ , \\
v^{t T}(r)&=&-C_3^{\rm IT}\left[C^{(2)}(r)-\frac{1}{r}\, C^{(1)}(r)\right] \ , \\
v^{b T}(r)&=& -C_4^{\rm IT}\frac{1}{r}C^{(1)}(r)\ ,\\
\end{eqnarray}
where, of course, only the $T\,$=$\,1$ component of the cutoff function enters in the CD interactions.
For ease of presentation, we have singled out the LO terms which only act in $S/T$=0/1 and 1/0
channels.  They are written as
\begin{eqnarray}
 v^c_{\rm LO}(r)&=&\frac{3}{16}\big[ C_{01}\, C_1(r)+ C_{10}\, C_0(r)\big]  \ , \\
v^\tau_{\rm LO}(r)&=& \frac{1}{16}\big[ C_{01}\, C_1(r)-3\,C_{10}\, C_0(r)\big]\ , \\
 v^\sigma_{\rm LO}(r)&=&\frac{1}{16}\big[ -3\,C_{01}\, C_1(r)+ C_{10}\, C_0(r)\big]\ , \\
 v^{\sigma\tau}_{\rm LO}(r)&=& -\frac{1}{16}\big[ C_{01}\, C_1(r)+ C_{10}\, C_0(r)\big]\ ,
\end{eqnarray}
where there are only two independent LECs, $C_{01}$ and $C_{10}$,
and the cutoff functions $C_0(r)$ and $C_1(r)$ are as given in Eq.~(\ref{eq:e2.6}).

\section{Fitted LECs at NLO and N3LO}
\label{app:a2}
Fitted values of the LECs obtained with interactions a-d and o
at NLO and N3LO are reported in Tables~\ref{tab:tb2anew} and~\ref{tab:tb2b},
respectively.
\begin{table}[bth]
\caption{\label{tab:tb2anew}%
The NLO LECs determined by fitting the $np$ and $pp$ database up to 15 MeV laboratory energy,
and the deuteron binding energy. }
\begin{ruledtabular}
\begin{tabular}{lccccc}
Model & a & b & c & d& o \\
\hline\hline
$C_{01}$(fm$^2$) & --.511051122E+01 & --.515205193E+01 & --.524089036E+01 & --.534645335E+01 &--.514608170E+01  \\
$C_{10}$(fm$^2$)   &  --.422732988E+01&--.486213195E+01  & --.147490885E+01&--.442765927E+01  & --.564430900E+01\\
%
\hline
$C_1$(fm$^4$)  & --.112720036E+01 & --.182744818E+01 & --.412069851E+01 & --.483330518E+01 & --.938734989E+00\\
$C_2$(fm$^4$)  & 0.909366063E+00 & 0.114092429E+01 & 0.251441807E+01 & 0.143873251E+01 & 0.483260368E+00\\
 $C_3$(fm$^4$) & 0.477208278E--01 & 0.353463551E+00 & 0.131550606E+01 & 0.145157319E+01 & 0.404430893E+00\\
$C_4$(fm$^4$)  & --.475987004E+00 & --.249962307E+00 & --.137446534E+00 & 0.143861202E+01 & --.531440872E+00\\
  $C_5$(fm$^4$)& 0.494135315E--01 & --.582318500E--02 & 0.688507262E+00 & 0.347184150E--01 & --.302484884E+00\\
$C_6$(fm$^4$)  & --.846846770E+00 & --.100082249E+01 & --.180046641E+01 & --.125608697E+01 & --.621725001E+00\\
$C_7$(fm$^4$)  & --.155550814E+01 & --.138788868E+01 & --.150745124E+01 & --.153475063E+01 & --.136793827E+01\\
\hline
\hline
$C^{\rm IT}_0$(fm$^4$)  & 0.190747072E--01 & 0.242061782E--01 & 0.343911021E--01 & 0.488093390E--01 & 0.219960910E--01\\
\end{tabular}
\end{ruledtabular}
\end{table}

\begin{table}[bth]
\caption{\label{tab:tb2b}%
The N3LO LECs determined by fitting the $np$ and $pp$ database up to 25 MeV laboratory energy,
and the deuteron binding energy.}
\begin{ruledtabular}
\begin{tabular}{lccccc}
Model & a & b & c & d & o\\
\hline\hline
 $C_{01}$(fm$^2$)  &--.511424764E+01  &--.508230349E+01  & --.503452047E+01 & --.503178655E+01 & --.512268575E+01\\
 $C_{10}$(fm$^2$)  &--.425743601E+01  & --.473602278E+01  &--.822959678E+00   & --.376510267E+01 &--.569749961E+01 \\
 %
 \hline
 $C_1$(fm$^4$)  & --.116115293E+01 &--.185257187E+01 & --.566863913E+01 & --.777131493E+01 & --.953469705E+00\\
 $C_2$(fm$^4$) & 0.903010818E+00 & 0.115714920E+01 & 0.193156537E+01 & 0.144559465E+01 & 0.475392426E+00\\
 $C_3$(fm$^4$)  & 0.306393229E--01 & 0.326127057E+00 & 0.614523475E+00 & 0.884483425E+00 & 0.399019793E+00\\
$C_4$(fm$^4$)  & --.482727638E+00 & --.255746003E+00 & --.401850350E+00 & 0.973261500E--01 & --.535686282E+00\\
 $C_5$(fm$^4$) & 0.609530095E--01 & --.121452200E--02 & 0.828348880E+00 & 0.922678262E+00 & --.324751963E+00\\
  $C_6$(fm$^4$)& --.849436923E+00 & --.995826514E+00 & --.141944181E+01 & --.505486549E+00 & --.648172781E+00\\
$C_7$(fm$^4$)  & --.148260980E+01 & --.121337175E+01 & --.149303317E+01 & --.175175554E+01 & --.134307736E+01\\
  \hline
 $D_1$(fm$^6$)  & 0.362483079E--02 & 0.135401332E+00 & --.101959190E+01 & --.119592294E+02 & 0.376343898E--01\\
$D_2$(fm$^6$)  & --.282368870E--02 & 0.587827922E--01 & --.152848879E+00 & --.314187328E+01 & 0.185768643E--01\\
$D_3$(fm$^6$)  & 0.655194092E--02 & 0.103792048E+00 & 0.172566911E+00 & --.159588592E+01 & 0.208685298E--01\\
$D_4$(fm$^6$)  & --.118076883E--02 & 0.281367204E--01 & 0.123221687E--01 & --.110427750E+01 & 0.104907424E--01\\
$D_5$(fm$^6$)  & 0.267103476E--01 & 0.466004385E--01 & --.100566414E+01 & --.283555072E+01 & 0.782433560E--02\\
$D_6$(fm$^6$)  & 0.131017125E--01 & 0.101301826E--01 & --.651798675E+00 & --.141483053E+01 & 0.189873465E--01\\
$D_7$(fm$^6$)  & 0.145918221E--01 & --.512549289E--01 & --.554593690E+00 & --.489668832E+01 & --.222332010E--01\\
$D_8$(fm$^6$)  & 0.112176977E--01 & --.427371460E--01 & 0.329865430E+00 & --.135461398E+01 & --.146786284E--01\\
$D_9$(fm$^6$)  & 0.111357163E+00 & 0.112830624E+00 & --.145021865E+00 & 0.564521782E+01 & 0.226506657E--01\\
$D_{10}$(fm$^6$)  & 0.325969493E--01 & 0.103419159E+00 & 0.990420811E+00 & --.354055305E+00 & 0.218482111E--01\\
$D_{11}$(fm$^6$)  & --.555205059E--01 & --.962540971E--01 & 0.341499700E--02 & 0.382300487E+01 & 0.936405658E--02\\
\hline\hline
$C^{\rm IT}_0$(fm$^4$)   & 0.616726547E--02 & --.221853840E--01 & --.406049402E--01 & --.127554450E+00 & 0.713292586E--02\\
 \hline
 $C^{\rm IT}_1$(fm$^6$)  & --.461868573E--02 & 0.852145060E--02 & 0.930442901E+00 & 0.236056581E+01 & --.113805789E--01\\
$C^{\rm IT}_2$(fm$^6$)  & --.806447857E--02 & --.276140270E--01 & 0.243860719E+00 & 0.572769698E+00 & --.126174063E--01\\
$C^{\rm IT}_3$(fm$^6$)   & --.261236310E--01 & --.156366250E--01 & --.372938280E--01 & --.299306179E--01 & 0.374105167E--03\\
$C^{\rm IT}_4$(fm$^6$)  & 0.156812161E--02 & 0.583713002E--01 & 0.137474019E+00 & --.218783861E+00 & 0.298742271E--01\\ 
\end{tabular}
\end{ruledtabular}
\end{table}


\begin{thebibliography}{100}
%
\bibitem{Weinberg:1990rz}
S.~Weinberg,
Phys.\ Lett.\ B \textbf{251}, 288-292 (1990)
doi:10.1016/0370-2693(90)90938-3

\bibitem{Weinberg:1991um}
S.~Weinberg,
Nucl.\ Phys.\ B \textbf{363}, 3-18 (1991)
doi:10.1016/0550-3213(91)90231-L

\bibitem{Ordonez:1992xp}
C.~Ordonez and U.~van Kolck,
Phys.\ Lett.\ B \textbf{291}, 459-464 (1992)
doi:10.1016/0370-2693(92)91404-W

\bibitem{Ordonez:1993tn}
C.~Ordonez, L.~Ray and U.~van Kolck,
Phys.\ Rev.\ Lett.\  \textbf{72}, 1982-1985 (1994)
doi:10.1103/PhysRevLett.72.1982

\bibitem{vanKolck:1994yi}
U.~van Kolck,
Phys.\ Rev.\ C \textbf{49}, 2932-2941 (1994)
doi:10.1103/PhysRevC.49.2932

\bibitem{Ordonez:1995rz}
C.~Ordonez, L.~Ray and U.~van Kolck,
Phys.\ Rev.\ C \textbf{53}, 2086-2105 (1996)
doi:10.1103/PhysRevC.53.2086
[arXiv:hep-ph/9511380 [hep-ph]].

\bibitem{Entem:2003ft}
D.~R.~Entem and R.~Machleidt,
Phys. Rev. C \textbf{68}, 041001 (2003)
doi:10.1103/PhysRevC.68.041001
[arXiv:nucl-th/0304018 [nucl-th]].

\bibitem{Epelbaum:2004fk}
E.~Epelbaum, W.~Glockle and U.~G.~Meissner,
Nucl. Phys. A \textbf{747}, 362-424 (2005)
doi:10.1016/j.nuclphysa.2004.09.107
[arXiv:nucl-th/0405048 [nucl-th]].

\bibitem{Epelbaum:2014efa}
E.~Epelbaum, H.~Krebs and U.~G.~Meißner,
Eur. Phys. J. A \textbf{51}, no.5, 53 (2015)
doi:10.1140/epja/i2015-15053-8
[arXiv:1412.0142 [nucl-th]].

\bibitem{Epelbaum:2014sza}
E.~Epelbaum, H.~Krebs and U.~Meissner,
Phys.\ Rev.\ Lett.\  \textbf{115}, no.12, 122301 (2015)
doi:10.1103/PhysRevLett.115.122301
[arXiv:1412.4623 [nucl-th]].


\bibitem{Piarulli:2014bda}
M.~Piarulli, L.~Girlanda, R.~Schiavilla, R.~Navarro Pérez, J.~E.~Amaro and E.~Ruiz Arriola,
Phys. Rev. C \textbf{91}, no.2, 024003 (2015)
doi:10.1103/PhysRevC.91.024003
[arXiv:1412.6446 [nucl-th]].



\bibitem{Entem:2017gor}
D.~R.~Entem, R.~Machleidt and Y.~Nosyk,
Phys. Rev. C \textbf{96}, no.2, 024004 (2017)
doi:10.1103/PhysRevC.96.024004
[arXiv:1703.05454 [nucl-th]].

\bibitem{Reinert:2017usi}
P.~Reinert, H.~Krebs and E.~Epelbaum,
Eur.\ Phys.\ J.\ A \textbf{54}, no.5, 86 (2018)
doi:10.1140/epja/i2018-12516-4
[arXiv:1711.08821 [nucl-th]].

\bibitem{Stoks:1994wp}
V.~G.~J.~Stoks, R.~A.~M.~Klomp, C.~P.~F.~Terheggen and J.~J.~de Swart,
Phys. Rev. C \textbf{49}, 2950-2962 (1994)
doi:10.1103/PhysRevC.49.2950
[arXiv:nucl-th/9406039 [nucl-th]].


\bibitem{Wiringa:1994wb}
R.~B.~Wiringa, V.~G.~J.~Stoks and R.~Schiavilla,
Phys. Rev. C \textbf{51}, 38-51 (1995)
doi:10.1103/PhysRevC.51.38
[arXiv:nucl-th/9408016 [nucl-th]].

\bibitem{Machleidt:2000ge}
R.~Machleidt,
Phys. Rev. C \textbf{63}, 024001 (2001)
doi:10.1103/PhysRevC.63.024001
[arXiv:nucl-th/0006014 [nucl-th]].

\bibitem{Chen:1999tn}
J.~W.~Chen, G.~Rupak and M.~J.~Savage,
Nucl. Phys. A \textbf{653}, 386-412 (1999)
doi:10.1016/S0375-9474(99)00298-5
[arXiv:nucl-th/9902056 [nucl-th]].

\bibitem{Bedaque:2002mn}
P.~F.~Bedaque and U.~van Kolck,
Ann. Rev. Nucl. Part. Sci. \textbf{52}, 339-396 (2002)
doi:10.1146/annurev.nucl.52.050102.090637
[arXiv:nucl-th/0203055 [nucl-th]].


\bibitem{Bethe:1949yr}
H.~A.~Bethe,
Phys. Rev. \textbf{76}, 38-50 (1949)
doi:10.1103/PhysRev.76.38

\bibitem{Kaplan:1998tg}
D.~B.~Kaplan, M.~J.~Savage and M.~B.~Wise,
Phys. Lett. B \textbf{424}, 390-396 (1998)
doi:10.1016/S0370-2693(98)00210-X
[arXiv:nucl-th/9801034 [nucl-th]].

\bibitem{Kaplan:1998we}
D.~B.~Kaplan, M.~J.~Savage and M.~B.~Wise,
Nucl. Phys. B \textbf{534}, 329-355 (1998)
doi:10.1016/S0550-3213(98)00440-4
[arXiv:nucl-th/9802075 [nucl-th]].

\bibitem{hammer:2020}
H.-W. Hammer, S. K\"onig, and U. van Kolck,
Rev. Mod. Phys. 92, 25004 (2020)
doi:10.1103/RevModPhys.92.025004
%
\bibitem{Konig:2016utl}
S.~König, H.~W.~Grießhammer, H.~W.~Hammer and U.~van Kolck,
Phys. Rev. Lett. \textbf{118}, no.20, 202501 (2017)
doi:10.1103/PhysRevLett.118.202501
[arXiv:1607.04623 [nucl-th]].


\bibitem{Gattobigio:2019omi}
M.~Gattobigio, A.~Kievsky and M.~Viviani,
Phys. Rev. C \textbf{100}, no.3, 034004 (2019)
doi:10.1103/PhysRevC.100.034004
[arXiv:1903.08900 [nucl-th]].

\bibitem{Bedaque:1998kg}
P.~F.~Bedaque, H.~W.~Hammer and U.~van Kolck,
Phys. Rev. Lett. \textbf{82}, 463-467 (1999)
doi:10.1103/PhysRevLett.82.463
[arXiv:nucl-th/9809025 [nucl-th]].

\bibitem{Bedaque:1998km}
P.~F.~Bedaque, H.~W.~Hammer and U.~van Kolck,
Nucl. Phys. A \textbf{646}, 444-466 (1999)
doi:10.1016/S0375-9474(98)00650-2
[arXiv:nucl-th/9811046 [nucl-th]].

\bibitem{Braaten:2004rn}
E.~Braaten and H.~W.~Hammer,
Phys. Rept. \textbf{428}, 259-390 (2006)
doi:10.1016/j.physrep.2006.03.001
[arXiv:cond-mat/0410417 [cond-mat]].

%
\bibitem{deltuva:2020}
A. Deltuva, M. Gattobigio, A. Kievsky, and M. Viviani,
Phys. Rev. C \textbf{102}, 064001 (2020)
%

\bibitem{Efimov:1970zz}
V.~Efimov,
Phys. Lett. B \textbf{33}, 563-564 (1970)
doi:10.1016/0370-2693(70)90349-7

\bibitem{Efimov:1971zz}
V.~N.~Efimov,
Sov. J. Nucl. Phys. \textbf{12}, 589 (1971)

\bibitem{Naidon:2016dpf}
P.~Naidon and S.~Endo,
Rept. Prog. Phys. \textbf{80}, no.5, 056001 (2017)
doi:10.1088/1361-6633/aa50e8
[arXiv:1610.09805 [quant-ph]].

\bibitem{Kievsky:2016kzb}
A.~Kievsky, M.~Viviani, M.~Gattobigio and L.~Girlanda,
Phys. Rev. C \textbf{95}, no.2, 024001 (2017)
doi:10.1103/PhysRevC.95.024001
[arXiv:1610.09858 [nucl-th]].

\bibitem{Epelbaum:2017byx}
E.~Epelbaum, J.~Gegelia and U.~G.~Meißner,
Nucl. Phys. B \textbf{925}, 161-185 (2017)
doi:10.1016/j.nuclphysb.2017.10.008
[arXiv:1705.02524 [nucl-th]].

\bibitem{Manohar:1983md}
A.~Manohar and H.~Georgi,
Nucl. Phys. B \textbf{234}, 189-212 (1984)
doi:10.1016/0550-3213(84)90231-1

\bibitem{Georgi:1992dw}
H.~Georgi,
Phys. Lett. B \textbf{298}, 187-189 (1993)
doi:10.1016/0370-2693(93)91728-6
[arXiv:hep-ph/9207278 [hep-ph]].

\bibitem{Kirscher:2009aj}
J.~Kirscher, H.~W.~Griesshammer, D.~Shukla and H.~M.~Hofmann,
Eur. Phys. J. A \textbf{44}, 239-256 (2010)
doi:10.1140/epja/i2010-10939-5
[arXiv:0903.5538 [nucl-th]].

\bibitem{Lensky:2016djr}
V.~Lensky, M.~C.~Birse and N.~R.~Walet,
Phys. Rev. C \textbf{94}, no.3, 034003 (2016)
doi:10.1103/PhysRevC.94.034003
[arXiv:1605.03898 [nucl-th]].

\bibitem{Epelbaum:2017tzp}
E.~Epelbaum, J.~Gegelia and U.~G.~Meißner,
Commun. Theor. Phys. \textbf{69}, no.3, 303 (2018)
doi:10.1088/0253-6102/69/3/303
[arXiv:1710.04178 [nucl-th]].

\bibitem{Piarulli:2016}
M.\ Piarulli, L.\ Girlanda, R.\ Schiavilla, A.\ Kievsky, A.\ Lovato, L.E.\ Marcucci, S.C.\ Pieper, M.\ Viviani, and R.B.\ Wiringa,
Phys.\ Rev.\ C {\bf 94}, 054007 (2016).
%

\bibitem{Girlanda:2020}
L.~Girlanda, A.~Kievsky, L.~E.~Marcucci and M.~Viviani,
Phys. Rev. C \textbf{102}, 064003 (2020).

\bibitem{Navarro:2013}
R.\ Navarro P\'erez, J.E.\ Amaro, and E.\ Ruiz Arriola,
Phys.\ Rev.\ C {\bf 88}, 064002 (2013).
%
\bibitem{Navarro:2014}
R.\ Navarro P\'erez, J.E.\ Amaro, and E.\ Ruiz Arriola,
Phys.\ Rev.\ C {\bf 89},  024004 (2014).
%
\bibitem{Navarro:2014b}
R.\ Navarro P\'erez, J.E.\ Amaro, and E.\ Ruiz Arriola,
Phys.\ Rev.\ C {\bf 89}, 064006 (2014).
%
\bibitem{Kortelainen:2010} 
M.\ Kortelainen, T.\ Lesinski, J.\ Mor\'{e}, W.\ Nazarewicz, J.\ Sarich, N.\ Schunck, M.V.\ Stoitsov, and S.\ Wild,
Phys.\ Rev.\ C {\bf 82}, 024313 (2010).
%
\bibitem{Bergervoet:1988}
J.\ R.\ Bergervoet, P.\ C.\ van Campen, W.\ A.\ van der Sanden, and J.\ J.\ de Swart, 
Phys. Rev. C {\bf 38} (1988) 15.
%
\bibitem{Sanden:1983}
W.\ A.\ van der Sanden, A.\ H.\ Emmen, and J.\ J.\ de Swart, 
Report No. THEF-NYM-83.11, Nijmegen (1983), unpublished; quoted in~\cite{Bergervoet:1988}.
%
\bibitem{Chen:2008}
Q.\ Chen {\it et al}., 
Phys.\ Rev. {\bf C} 77 (2008) 054002.
%
\bibitem{Miller:1990}
G.\ A.\ Miller, M.\ K.\ Nefkens, and I.\ Slaus, 
Phys.\ Rep. {\bf 194} (1990) 1.
%
\bibitem{Ericson:1983}
T.\ E.\ O.\ Ericson and M.\ Rosa-Clot, 
Nucl.\ Phys.\ A {\bf 405}, 497 (1983).
%
\bibitem{Rodning:1990}
N.\ L.\ Rodning and L.\ D.\ Knutson, 
Phys.\ Rev.\ C {\bf 41}, 898 (1990).
%
\bibitem{Huber:1998}
A.\ Huber, T.\ Udem, B.\ Gross, J.\ Reichert, M.\ Kourogi, K.\ Pachucki, M.\ Weitz, and T.\ W.\ Hansch, 
Phys. Rev. Lett. {\bf 80}, 468 (1998).
%
\bibitem{Martorell:1995}
J.\ Martorell, D.\ W.\ L.\ Sprung, and D.\ C.\ Zheng, 
Phys.\ Rev.\ C {\bf 51}, 1127 (1995).
%
\bibitem{Stoks:1993}
V.G.J.\ Stoks, R.A.M.\ Klomp, M.C.M.\ Rentmeester, and J.J.\ de Swart,
Phys.\ Rev.\ C {\bf 48}, 792 (1993).
%
\bibitem{Gross:2008}
F.L.\ Gross and A.\ Stadler,
Phys.\ Rev.\ C {\bf 78}, 014005 (2008).
%
\bibitem{Kievsky:2008es}
A.~Kievsky, S.~Rosati, M.~Viviani, L.~E.~Marcucci and L.~Girlanda,
J. Phys. G \textbf{35}, 063101 (2008)
doi:10.1088/0954-3899/35/6/063101
[arXiv:0805.4688 [nucl-th]].

\bibitem{Marcucci:2019hml}
L.~E.~Marcucci, J.~Dohet-Eraly, L.~Girlanda, A.~Gnech, A.~Kievsky and M.~Viviani,
Front. in Phys. \textbf{8}, 69 (2020)
doi:10.3389/fphy.2020.00069
[arXiv:1912.09751 [nucl-th]].

\bibitem{Thomas:1935zz}
L.~H.~Thomas,
Phys. Rev. \textbf{47}, 903-909 (1935)
doi:10.1103/PhysRev.47.903

\bibitem{Girlanda:2011fh}
L.~Girlanda, A.~Kievsky and M.~Viviani,
Phys. Rev. C \textbf{84}, no.1, 014001 (2011)
doi:10.1103/PhysRevC.84.014001. Erratum  Phys.\ Rev.\ C \textbf{102}, 019903 (2020) doi:10.1103/PhysRevC.84.014001 [arXiv:1102.4799 [nucl-th]].

\bibitem{Girlanda:2018xrw}
L.~Girlanda, A.~Kievsky, M.~Viviani and L.~E.~Marcucci,
Phys. Rev. C \textbf{99}, no.5, 054003 (2019)
doi:10.1103/PhysRevC.99.054003
[arXiv:1811.09398 [nucl-th]].


\bibitem{Kievsky:2015dtk}
A.~Kievsky and M.~Gattobigio,
Few Body Syst. \textbf{57}, no.3, 217-227 (2016)
doi:10.1007/s00601-016-1049-5
[arXiv:1511.09184 [nucl-th]].

\bibitem{Kievsky:2018xsl}
A.~Kievsky, M.~Viviani, D.~Logoteta, I.~Bombaci and L.~Girlanda,
Phys. Rev. Lett. \textbf{121}, no.7, 072701 (2018)
doi:10.1103/PhysRevLett.121.072701
[arXiv:1806.02636 [nucl-th]].

\bibitem{Gnech:2020qtt}
A.\ Gnech, M.\ Viviani, and L.E.\ Marcucci,
Phys.\ Rev.\ C {\bf 102}, 014001 (2020)
doi:10.1103/PhysRevC.102.014001
[arXiv:2004.05814 [nucl-th]].
%
\bibitem{Gnech:2021}
A.\ Gnech to be published.
%
\bibitem{kievsky:2017}
A.\ Kievsky, A.\ Polls, B.\ Juli\'a D\'\i az, and N.K.\ Timofeyuk,
Phys.\ Rev.\ C {\bf 96}, 040501(R) (2017)
%
\bibitem{kievsky:2020}
A.\ Kievsky, A.\ Polls, B.\ Juli\'a D\'\i az, N.K.\ Timofeyuk, and M.\ Gattobigio,
Phys.\ Rev.\ C {\bf 102}, 063320 (2020)
%
\bibitem{Schmidt99}
K.E.\ Schmidt and S.\ Fantoni,
Phys.\ Lett.\ B {\bf 446}, 99 (1999).
%
\bibitem{Gandolfi20}
S.\ Gandolfi, D.\ Lonardoni, A.\ Lovato, and M.\ Piarulli,
Front.\ in Phys.\ {\bf 8}, 117 (2020).
%
\bibitem{Lonardoni18}
D.\ Lonardoni, J.\ Carlson, S.\ Gandolfi, J.\ E.\ Lynn, K.\ E.\ Schmidt, A.\ Schwenk, and X.\ Wang,
Phys.\ Rev. Lett.\ {\bf 120}, 122502 (2018)
%
\bibitem{Lonardoni18b}
D.\ Lonardoni, S.\ Gandolfi, J.\ E.\ Lynn, C.\ Petrie, J.\ Carlson, K.\ E.\ Schmidt, and A.\ Schwenk,
Phys.\ Rev.\ C {\bf 97}, 044318 (2018)
%
\bibitem{Gandolfi13}
S.\ Gandolfi, A.\ Lovato, J.\ Carlson, and K.\ E.\ Schmidt,
Phys.\ Rev.\ C {\bf 90}, 061306 (2014)
%
\bibitem{Contessi17}
L.\ Contessi, A.\ Lovato, F.\ Pederiva, A.\ Roggero, J.\ Kirscher, and U.~van Kolck,
Phys.\ Lett.\ B {\bf 772}, 839 (2017)
%
\bibitem{Ekstrom15}
A.\ Ekstr\"{o}m, G.R.\ Jansen, K.A.\ Wendt, G.\ Hagen, T.\ Papenbrock, B.D.\ Carlsson, C.\ Forss\'{e}n, M.\ Hjorth-Jensen, P.\ Navr\'{a}til, and W.\ Nazarewicz,
Phys. Rev. C 91, 051301(R) (2015).
\end{thebibliography}
\end{document}